\documentclass[onecolumn,eqsecnum,pra]{revtex4}

\usepackage{amssymb}
\usepackage{amsmath}

\begin{document}

\title{Measured quantum probability distribution functions for Brownian
motion}

\author{G. W. Ford}
\affiliation{Department of Physics, University of Michigan, Ann Arbor, MI
48109-1040}

\author{R. F. O'Connell}
\affiliation{Department of Physics and Astronomy, Louisiana State
University, Baton Rouge, LA 70803-4001}

\date{\today }

\begin{abstract}
The quantum analog of the joint probability distributions describing a
classical stochastic process is introduced. A prescription is given for
constructing the quantum distribution associated with a sequence of
measurements. For the case of quantum Brownian motion this prescription is
illustrated with a number of explicit examples. In particular it is shown
how the prescription can be extended in the form of a general formula for
the Wigner function of a Brownian particle entangled with a heat bath..
\end{abstract}


\maketitle

\section{Introduction}

\label{sec:one}

The notion of joint probability distribution is basic to the description of
classical stochastic processes. The purpose here is to describe the
extension to the quantum regime, giving the prescription for constructing
the quantum joint probability distribution associated with a sequence of
measurements. The prescription is illustrated with a number of examples.
These are an important part of this work, since they show how the
prescription can be used to calculate a variety of quantities of practical
interest.

The idea is simple: The system is initially in thermal equilibrium. A first
measurement prepares a state that then develops in time according to the
underlying dynamics. Then a second measurement prepares a new state. And so
on until the final measurement in the sequence.

In this connection it is necessary to consider the description of quantum
measurement. In its most naive form, found in many textbooks, a quantum
measurement of a dynamical variable is described as a projection of the
system into an eigenstate of the variable, with no memory of the previous
state. In particular for a variable with a continuous spectrum, such as the
position of a Brownian particle, with no square integrable eigenstate, this
naive description is unsatisfactory. In an earlier publication jointly with
J. T. Lewis \cite{ford86}, the description of measurement as applied to
quantum stochastic processes was addressed in some detail. Since that
earlier publication may not be accessible to all readers, in Section \ref
{sec:three} a summary is given of the essential features of quantum
measurement as they apply to the definition of the distribution functions.
The reader will observe that the description given there involves no new
theory of quantum measurement. Rather, a prescription is adopted based on
that used by many authors making practical calculations related to real
experiments.

The plan of the paper is as follows. To begin, in a brief Section \ref
{sec:two} the joint distribution functions of classical mechanics are
described. The quantum joint distribution functions, which are a close
analog of the classical quantities, are introduced in Section \ref{sec:three}
. There the key result is the prescription (\ref{3.29}) for the joint
distribution function associated with $n$ successive measurements. For the
case of quantum Brownian motion this prescription can be readily evaluated
to give explicit closed form expressions. Therefore in Appendix \ref
{appendix:A} a review is given of those aspects of the theory of quantum
Brownian motion that will be useful in the applications. There the key
quantities needed for the later discussion are the commutator and the mean
square displacement, given by the general expressions (\ref{A.7}) and (\ref
{A.11}). Later in Appendix \ref{appendix:A} these expressions are evaluated
explicitly for the Ohmic and single relaxation time models, and the results
compared with approximate expressions obtained by master equation methods.
In Section \ref{sec:four} the results of Appendix \ref{appendix:A} are used
to evaluate the characteristic function associated with the distribution
function describing $n$ successive measurements. There the important result
is the expression (\ref{4.2}) for the characteristic function, where it is
seen explicitly how in classical mechanics, where the commutator vanishes,
the effects of measurement can be separated from the underlying stochastic
process. As an application of this result, an explicit expression is
constructed for the pair distribution function associated with wave packet
spreading. In Section \ref{sec:five} we discuss the probability
distribution, which corresponds to what in elementary quantum mechanics is
\textquotedblleft the square of the wave function\textquotedblright . A
general expression is obtained that is illustrated first with the example of
wave packet spreading and then with the example of a \textquotedblleft Schr
\"{o}dinger cat\textquotedblright\ state. In either case the discussion
includes the case of a free particle as well as that of a particle in a
harmonic well. Finally, in Section \ref{sec:six} the Wigner function is
introduced. There the key result is the simple formula (\ref{6.5}) for the
Wigner characteristic function (the Fourier transform of the Wigner
function). The Wigner function corresponds to the phase space distribution
of classical mechanics, but is definitely not a probability distribution (it
is seen explicitly in the examples that the Wigner function need not be
positive) and cannot be the result of direct quantum measurement.
Nevertheless, the Wigner function is useful for describing the results of
measurement. In particular the probability distribution in coordinate or
momentum are obtained by integration over the conjugate variable.

\section{Distribution functions in classical Brownian motion}

\label{sec:two}

In the theory of classical Brownian motion a stochastic process is
completely described by a hierarchy of probability distributions. Here the
standard reference for the physicist is the review article by Wang and
Uhlenbeck \cite{wang}, reprinted in the \textquotedblleft
Noisebook\textquotedblright\ \cite{wax}. For a stochastic variable $y(t)$
one introduces
\begin{eqnarray}
W(y_{1},t_{1})dy_{1} &=&\text{probability of finding }y(t_{1})\text{ in the
interval }dy_{1}\text{ about }y_{1},  \notag \\
W(y_{1},t_{1};y_{2},t_{2})dy_{1}dy_{2} &=&\text{probability of finding }
y(t_{1})\text{ in the interval }dy_{1}\text{ about }y_{1}  \notag \\
&&\text{\emph{and} }y(t_{2})\text{ in the interval }dy_{2}\text{ about }
y_{2},  \notag \\
&&\text{and so on.}  \label{2.1}
\end{eqnarray}
This hierarchy must satisfy the following more or less obvious conditions

\begin{quotation}
1. Positivity

$\qquad W(y_{1},t_{1};y_{2},t_{2};\cdots ;y_{n},t_{n})\geq 0.$

2. Symmetry

$\qquad W(y_{1},t_{1};y_{2},t_{2};\cdots ;y_{n},t_{n})$ is a symmetric
function of the set of variables $y_{1},t_{1};y_{2},t_{2},\cdots
,y_{n},t_{n} $.

3. Consistency

$\qquad W(y_{1},t_{1};y_{2},t_{2};\cdots ;y_{n},t_{n})=\int
dy_{n+1}W(y_{1},t_{1};y_{2},t_{2};\cdots ;y_{n},t_{n};y_{n+1},t_{n+1})$.
\end{quotation}

\noindent Note that consistency corresponds to conservation of total
probability,
\begin{equation}
1=\int dy_{1}\int dy_{2}\cdots \int dy_{n}W(y_{1},t_{1};y_{2},t_{2};\cdots
;y_{n},t_{n}).  \label{2.2}
\end{equation}
A theorem of Kolomogorov states that if the hierarchy satisfies these
conditions there must exist an underlying classical process \cite
{kolmogorov56}. That is, there must exist an ensemble of time-tracks $y(t)$
such that the $W$'s are the weighted fraction of time tracks that go through
the appropriate intervals.

A natural question is How is this description changed in the quantum case?
The answer will be seen in the following Section, but for now one can say
that the essential change is that in the quantum case the symmetry condition
no longer holds.

\section{Quantum Distribution functions}

\label{sec:three}

The system considered is that of a Brownian particle coupled to a heat bath,
a system with an infinite number of degrees of freedom. The quantum \
mechanical motion of this system is described by a microscopic Hamiltonian $
H $ and corresponds to a unitary transformation of states in Hilbert space, 
\begin{equation}
\Psi (t)=U(t)\Psi (0),  \label{3.1}
\end{equation}
where $\Psi (t)$ is the state vector at time $t$ and 
\begin{equation}
U(t)=\exp \{-iHt/\hbar \}.  \label{3.2}
\end{equation}
However, one does not have precise knowledge of the initial state. Instead,
there is an initial density matrix. The density matrix is defined as an
operator $\rho (t)$ in Hilbert space such that $\left\langle \Phi ,\rho
(t)\Phi \right\rangle /\left\langle \Phi ,\Phi \right\rangle $ is the
relative probability at time $t$ that the system is in any given state $\Phi 
$ \cite{dirac47,fano57}. Note that consistent with this definition $\rho $
must be a positive definite Hermitian operator. Its time development follows
from (\ref{3.1}), 
\begin{equation}
\rho (t)=U(t)\rho (0)U^{\dag }(t).  \label{3.3}
\end{equation}

Before introducing the distribution functions, it is necessary to make some
general remarks about measurement in quantum mechanics. By \textquotedblleft
measurement\textquotedblright\ here is meant \textquotedblleft measurement
with selection\textquotedblright\ (what Pauli in his famous Ziffer 9 called
\textquotedblleft measurement of the second kind\textquotedblright\ \cite
{pauli}) so that measurement irreversibly changes the state of the system.
When discussed in general terms in textbooks, the accepted description of
this change of state is framed in terms of measurement of a discrete
variable (Hermitian operator with a pure point spectrum). Let $B$ be such a
variable, with $b$ an eigenvalue and $P_{b}$ its associated projection
operator, so that 
\begin{equation}
B=\sum_{b}bP_{b}.  \label{3.4}
\end{equation}
Then the effect of a measurement at time $t_{1}$ whose result is that the
eigenvalue $b$ is in the interval $M$ is to instantaneously transform the
density matrix, 
\begin{equation}
\rho (t_{1})\rightarrow P_{M}\rho (t_{1})P_{M},  \label{3.5}
\end{equation}
where $P_{M}$ is the projection operator associated with the interval, 
\begin{equation}
P_{M}=\sum_{b\in M}P_{b}.  \label{3.6}
\end{equation}
Here instantaneous means that the duration of the measurement is short
compared with the natural periods of the system. The prescription (\ref{3.5}
) may be obtained from various assumptions about the optimal character of
the measurement (such that the disturbance of the state is somehow minimal).
See, e.g., L\"{u}ders \cite{luders51}, Goldberger and Watson \cite
{goldberger64}, Furry \cite{furry66}, or Davies and Lewis \cite{davies70}.

But the prescription (\ref{3.5}) is too restricted for our purpose; it
represents too limited a class of measurements. We must consider
measurements of limited precision and involving operators with a continuous
spectrum, such as the position of a Brownian particle. For guidance as to
how to generalize the prescription, we look to such practical fields as the
theory of angular correlations (see, e. g., the article by Frauenfelder and
Steffan \cite{siegbahn_vol2} \S 3) or the theory of polarization in multiple
scattering (see, e. g., Wolfenstein \cite{wolfenstein56} \S 4). There one
associates a transition operator $T$ (in the scattering case this would be
the Wigner T-matrix) with a measurement with a given result (e.g., the
observation of an emitted gamma ray in a given direction) and represents the
transformation of $\rho $ brought about by the measurement at time $t_{1}$
by 
\begin{equation}
\rho (t_{1})\rightarrow T\rho (t_{1})T^{\dag }.  \label{3.7}
\end{equation}
Note that this is the most general transformation that preserves the
positivity of the density matrix. In the special case where the transition
operator is a projection operator, one recovers the prescription (\ref{3.5}
). To be consistent with the probabilistic interpretation of the density
matrix, the transition operator must satisfy the general requirement 
\begin{equation}
\left\Vert T\right\Vert ^{2}\equiv \max_{\Phi }\frac{\left\langle T\Phi
,T\Phi \right\rangle }{\left\langle \Phi ,\Phi \right\rangle }\leq 1.
\label{3.8}
\end{equation}
The diagonal matrix elements of $\rho $ formed with respect to a complete
set of stated are interpreted as the probabilities of finding the system in
the corresponding states. The sum of these probabilities over all states is
the trace and, as a consequence of the requirement (\ref{3.8}), this is
reduced by measurement. Thus, 
\begin{equation}
\mathrm{Tr}\{T\rho T^{\dag }\}=\mathrm{Tr}\{\rho T^{\dag }T\}\leq \mathrm{Tr}
\{\rho \}\left\Vert T\right\Vert ^{2}\leq \mathrm{Tr}\{\rho \}.  \label{3.9}
\end{equation}
In fact the ratio $\mathrm{Tr}\{\rho T^{\dag }T\}/\mathrm{Tr}\{\rho \}$ can
be interpreted as the probability that measurement will produce the given
result.

This reduction of the sum over all states of the probability of finding the
system in each state is not a unique feature of quantum probability. In
classical probability, where the probabilities after measurement would be
interpreted as joint probabilities of the result of the measurement \emph{and
} of finding the system in the state, the sum of probabilities is also
reduced. The difference is that in the classical case none of the individual
probabilities will be increased, while in the quantum case some may increase.

As a simple example illustrating all this, consider a spin 1/2 system
initially polarized in the $+z$ direction. This can be represented by the $
2\times 2$ density matrix
\begin{equation}
\rho _{0}=\left( 
\begin{array}{cc}
1 & 0 \\ 
0 & 0
\end{array}
\right) .  \label{3.10}
\end{equation}
Note that the diagonal elements, which are, respectively, the probability
that the spin is in the $+z$ and $-z$ directions, add up to $1$. This is in
accord with the \emph{convention} that the trace of the density matrix is
normalized to $1$ prior to the first measurement. Suppose that a measurement
is made, for example by a Stern-Gerlach apparatus, whose result is that the
spin is in the $+x$ direction. The transition matrix corresponding to this
result is
\begin{equation}
T=\frac{1}{2}\left( 
\begin{array}{cc}
1 & 1 \\ 
1 & 1
\end{array}
\right) ,  \label{3.11}
\end{equation}
in this case a projection operator. The density matrix after the measurement
is
\begin{equation}
\rho _{+x}=T\rho _{0}T^{\dag }=\left( 
\begin{array}{cc}
\frac{1}{4} & \frac{1}{4} \\ 
\frac{1}{4} & \frac{1}{4}
\end{array}
\right) .  \label{3.12}
\end{equation}
The probability that the spin is in the $+z$ direction has decreased from $1$
to $\frac{1}{4}$ while the probability that the spin is in the $-z$
direction has \emph{increased} from $0$ to $\frac{1}{4}$ . Nevertheless the
sum of the probabilities is $\frac{1}{2}$, less than the sum before the
measurement, One might ask: what happened to the probability, how is it that
the probabilities after the measurement don't add up to $1$? The answer is
that there is another possible result of the measurement: the spin is in the 
$-x$ direction. Repeating the above argument for this case, we find for the
density matrix after the measurement
\begin{equation}
\rho _{-x}=\left( 
\begin{array}{cc}
\frac{1}{4} & -\frac{1}{4} \\ 
-\frac{1}{4} & \frac{1}{4}
\end{array}
\right) .  \label{3.13}
\end{equation}
\ Again the probability that the spin after the measurement is in the $\pm z$
direction is $\frac{1}{4}$. Thus, in either case the sum of the
probabilities is $\frac{1}{2}$, the probability that the measurement
produces the given result. These probabilities add up to $1$, so overall
probability is conserved.

With these remarks as a guide, consider the measurement of a dynamical
variable $y$. Assume that $y$ is a variable, such as the position or
velocity of the Brownian particle, with a continuous spectrum over all real
values. In this case one can associate with the measurement a function $
f(y_{1})$ such that (here $y_{1}$ is a c-number) 
\begin{equation}
\frac{\left\langle f(y-y_{1})\Phi ,f(y-y_{1})\Phi \right\rangle }{
\left\langle \Phi ,\Phi \right\rangle }dy_{1}  \label{3.14}
\end{equation}
is the conditional probability that if the system is in state $\Phi $ the
instrument will read in the interval $dy_{1}$ about $y_{1}$. Here the choice
that only the difference $y-y_{1}$ appears is made for convenience, since by
choosing $f(y_{1})$ to be peaked the requirement that the measured value be
somehow close to the actual value can be satisfied in an obvious way . If
this conditional probability is to be normalized, one must require ($f$ need
not be real) 
\begin{equation}
\int_{-\infty }^{\infty }dy_{1}\left\vert f(y_{1})\right\vert ^{2}=1.
\label{3.15}
\end{equation}
An example to keep in mind is that of a \textquotedblleft Gaussian
instrument\textquotedblright\ \cite{feynman}, for which 
\begin{equation}
f(y_{1})=\frac{1}{(2\pi \sigma _{1}^{2})^{1/4}}\exp \{-\frac{y_{1}^{2}}{
4\sigma _{1}^{2}}\},  \label{3.16}
\end{equation}
where $\sigma _{1}$ is the experimental width. The result of a measurement
at time $t_{1}$ in which the instrument reads in the interval $I_{1}$ is
therefore to instantaneously transform $\rho ,$
\begin{equation}
\rho (t_{1})\rightarrow \int_{I_{1}}dy_{1}f(y-y_{1})\rho
(t_{1})f(y-y_{1})^{\dag }.  \label{3.17}
\end{equation}
As remarked above, an instantaneous measurement is to be understood as one
whose duration is short compared with the natural periods of the motion.

The distribution functions are now constructed as follows. To begin assume
that at $t=0$ (or in the distant past) the system is in equilibrium at
temperature $T$, 
\begin{equation}
\rho (0)=\rho _{0}\equiv \frac{e^{-H/kT}}{\mathrm{Tr}\{e^{-H/kT}\}},
\label{3.18}
\end{equation}
so $\rho $ is initially normalized. If $y$ is to be measured at a later time 
$t_{1}$ the system must move in time from $0$ to $t_{1}$, 
\begin{equation}
\rho _{0}\rightarrow U(t_{1})\rho _{0}U^{\dag }(t_{1}),  \label{3.19}
\end{equation}
according to (\ref{3.3}). At $t_{1}$ a measurement is made, 
\begin{equation}
U(t_{1})\rho _{0}U^{\dag }(t_{1})\rightarrow
\int_{I_{1}}dy_{1}f(y-y_{1})U(t_{1})\rho _{0}U^{\dag
}(t_{1})f(y-y_{1})^{\dag },  \label{3.20}
\end{equation}
according to (\ref{3.17}). Finally, the system moves from $t_{1}$ to $t$, 
\begin{eqnarray}
&&\int_{I_{1}}dy_{1}f(y-y_{1})U(t_{1})\rho _{0}U^{\dag
}(t_{1})f(y-y_{1})^{\dag }  \notag \\
&\rightarrow &\int_{I_{1}}dy_{1}U(t-t_{1})f(y-y_{1})U(t_{1})\rho _{0}U^{\dag
}(t_{1})f(y-y_{1})^{\dag }U^{\dag }(t-t_{1}).  \label{3.21}
\end{eqnarray}
This last expression can be simplified somewhat by introducing the
time-dependent variable (Heisenberg representation), 
\begin{equation}
y(t_{1})=U^{\dag }(t_{1})yU(t_{1}).  \label{3.22}
\end{equation}
Noting that $U(t-t_{1})=U(t)U(-t_{1})=U(t)U^{\dag }(t_{1})$, the final
density matrix (\ref{3.21}) can be written 
\begin{equation}
\int_{I_{1}}dy_{1}U(t)f[y(t_{1})-y_{1}]\rho _{0}f[y(t_{1})-y_{1}]^{\dag
}U^{\dag }(t).  \label{3.23}
\end{equation}
The probability that the measurement of $y$ is in $I_{1}$is the trace of
this final density matrix. This same probability is interpreted as the
integral over the one-point distribution $W(y_{1},t_{1})$ over the interval $
I_{1}$, so that 
\begin{equation}
\int_{I_{1}}dy_{1}W(y_{1},t_{1})=\int_{I_{1}}dy_{1}\mathrm{Tr}
\{U(t)f[y(t_{1})-y_{1}]\rho _{0}f[y(t_{1})-y_{1}]^{\dag }U^{\dag }(t)\}.
\label{3.24}
\end{equation}
Since $I_{1}$ is arbitrary, one can identify 
\begin{equation}
W(y_{1},t_{1})=\mathrm{Tr}\{U(t)f[y(t_{1})-y_{1}]\rho
_{0}f[y(t_{1})-y_{1}]^{\dag }U^{\dag }(t)\}.  \label{3.25}
\end{equation}
Finally, the trace is invariant under cyclic permutation of the factors, so
one can write 
\begin{equation}
W(y_{1},t_{1})=\mathrm{Tr}\{f[y(t_{1})-y_{1}]\rho
_{0}f[y(t_{1})-y_{1}]^{\dag }\}.  \label{3.26}
\end{equation}

In the same way one can show that the two-point distribution is 
\begin{equation}
W(y_{1},t_{1};y_{2},t_{2})=\mathrm{Tr}\{f[y(t_{2})-y_{2}]f[y(t_{1})-y_{1}]
\rho _{0}f[y(t_{1})-y_{1}]^{\dag }f[y(t_{2})-y_{2}]^{\dag }\}  \label{3.27}
\end{equation}
and in general, using an obvious shorthand notation,
\begin{equation}
W(1,\cdots ,n)=\mathrm{Tr}\{f(n)\cdots f(1)\rho _{0}f(1)^{\dag }\cdots
f(n)^{\dag }\}.  \label{3.28}
\end{equation}
Finally, note that under cyclic permutation of the factors in the trace, one
can write the expression for the $n$-point distribution in the compact form:
\begin{equation}
W(1,\cdots ,n)=\left\langle f(1)^{\dag }\cdots f(n)^{\dag }f(n)\cdots
f(1)\right\rangle ,  \label{3.29}
\end{equation}
where the angular brackets indicate the thermal equilibrium expectation.
That is, for a given operator $\mathcal{O}$, 
\begin{equation}
\left\langle \mathcal{O}\right\rangle \equiv \mathrm{Tr}\{\mathcal{O}\rho
_{0}\}.  \label{3.30}
\end{equation}
The expression (\ref{3.29}) is the key result of this section.

In connection with the expression (\ref{3.29}) for the joint probability
distribution, it should first be emphasized that it is time-ordered: $
0<t_{1}<t_{2}<\cdots <t_{n}$. Also we should point out that in our shorthand
notation the label applies to \emph{all} the parameters of the measurement.
Thus, for example the label \textquotedblleft $j$\textquotedblright\
represents not only the value $y_{j}$ and the time $t_{j}$ but also the
instrumental parameters such as the width $\sigma _{j}$. The symmetry
property of the classical stochastic process refers to symmetry under
permutations of the labels. In other words, the probability distributions of
a classical stochastic process are symmetric under the interchange of all
the parameters of the measurements. An inspection of the expression (\ref
{3.29}) for the quantum probability distribution shows that it does not have
that symmetry, because the $f$'s at different times do not in general
commute. However the quantum distributions still have the consistency
property, providing the integral is over the results of the last
measurement. This is sometimes called marginal consistency.

\section{Characteristic functions}

\label{sec:four}

In this section we consider the evaluation of the distribution functions
when the dynamical variable $y(t)$ is taken to be the position operator $
x(t) $ for quantum Brownian motion, introduced in Appendix \ref{appendix:A}.
In evaluating these distribution functions, it is convenient in analogy with
the classical case to introduce the corresponding characteristic functions,
defined by
\begin{equation}
\xi (1,\cdots ,n)=\int_{-\infty }^{\infty }dx_{1}\cdots \int_{-\infty
}^{\infty }dx_{n}W(1,\cdots ,n)\exp \{i\sum_{j=1}^{n}k_{j}x_{j}\}.
\label{4.1}
\end{equation}
An important result is that for quantum Brownian motion the characteristic
function can be cast in the form
\begin{equation}
\xi (1,\cdots ,n)=K(1,\cdots ,n)\left\langle \exp
\{i\sum_{j=1}^{n}k_{j}x(t_{j})\}\right\rangle ,  \label{4.2}
\end{equation}
where the factor $K(1,\cdots ,n)$ is given by
\begin{equation}
K(1,\cdots ,n)=\prod_{j=1}^{n}\int_{-\infty }^{\infty }dx_{j}f^{\ast
}(x_{j}-\sum_{l=j+1}^{n}k_{l}\frac{[x(t_{j}),x(t_{l})]}{2i}
)f(x_{j}+\sum_{l=j+1}^{n}k_{l}\frac{[x(t_{j}),x(t_{l})]}{2i})e^{ik_{j}x_{j}}.
\label{4.3}
\end{equation}
In this expression the sums are to be taken to be zero when $j=n$. The
importance of this result lies first of all in the fact that the effects of
measurement are completely contained in the factor $K$, in which the
particle dynamics enters only through the commutators. In the classical
limit these commutators vanish and $K$ becomes a simple numerical factor.
Indeed in this classical limit one generally considers measurements of
perfect precision, for which $\left\vert f(x_{j})\right\vert ^{2}\rightarrow
\delta (x_{j})$ and the factor $K$ is unity. The expression (\ref{4.2}) then
becomes the familiar form for classical Brownian motion \cite{doob}. On the
other hand, in the quantum case, where the commutators do not vanish, \ it
is clear that the particle dynamics is inextricably linked with measurement
and, as a consequence, the symmetry property of classical stochastic
processes does not hold. A second reason for the importance of this result
is that it is convenient for calculation, as we shall illustrate in the
example below. Before \ that, however, we give a brief derivation.

Consider first the case $n=1$. Using the expression (\ref{3.29}) in the
definition (\ref{4.1}) of the characteristic function, we can write
\begin{equation}
\xi (1)=\left\langle \int_{-\infty }^{\infty }dx_{1}f^{\ast
}\{x_{1}-x(t_{1})\}f\{x_{1}-x(t_{1})\}e^{ik_{1}x_{1}}\right\rangle .
\label{4.4}
\end{equation}
Making the change of variable $x_{1}\rightarrow x_{1}+x(t_{1})$, we see that
\begin{equation}
\xi (1)=K(1)\left\langle e^{ik_{1}x(t_{1})}\right\rangle ,  \label{4.5}
\end{equation}
where
\begin{equation}
K(1)=\int_{-\infty }^{\infty }dx_{1}f^{\ast }(x_{1})f(x_{1})e^{ik_{1}x_{1}}.
\label{4.6}
\end{equation}
Recall that the sums in the expression (\ref{4.3}) are to be taken to be
zero when $j=n$.

Next consider
\begin{eqnarray}
\xi (1,2) &=&\left\langle \int_{-\infty }^{\infty }dx_{1}\int_{-\infty
}^{\infty }dx_{2}f(1)^{\dag }f(2)^{\dag
}f(2)f(1)e^{i(k_{1}x_{1}+k_{2}x_{2})}\right\rangle  \notag \\
&=&K(2)\left\langle \int_{-\infty }^{\infty }dx_{1}f^{\ast
}\{x_{1}-x(t_{1})\}e^{i\{k_{1}x_{1}+k_{2}x(t_{2})\}}f\{x_{1}-x(t_{1})\}
\right\rangle .  \label{4.7}
\end{eqnarray}
Here $K(2)$ is exactly of the form (\ref{4.6}) but with the label
\textquotedblleft $2$\textquotedblright\ in place of \textquotedblleft $1$
\textquotedblright . Remember also that with our shorthand notation the
label also represents the instrumental parameters, so the measurement
function $f$ changes with the label. Next, we apply the generalized
Baker-Campbell-Hausdorf formula (\ref{B.4}) to write
\begin{equation}
e^{i\{k_{1}x_{1}+k_{2}x(t_{2})\}}f\{x_{1}-x(t_{1})\}=f
\{x_{1}-x(t_{1})+ik_{2}[x(t_{1}),x(t_{2})]\}e^{i\{k_{1}x_{1}+k_{2}x(t_{2})
\}}.  \label{4.8}
\end{equation}
With this and making the change of variable $x_{1}\rightarrow
x_{1}+x(t_{1})+k_{2}[x(t_{1}),x(t_{2})]/2i$, we obtain
\begin{equation}
\xi (1,2)=K(1,2)\left\langle
e^{ik_{1}x(t_{1})}e^{ik_{2}x(t_{2})}e^{k_{1}k_{2}[x(t_{1}),x(t_{2})]}\right
\rangle ,  \label{4.9}
\end{equation}
where
\begin{equation}
K(1,2)=\int_{-\infty }^{\infty }dx_{1}f^{\ast }\{x_{1}-k_{2}\frac{
[x(t_{1}),x(t_{2})]}{2i}\}f\{x_{1}+k_{2}\frac{[x(t_{1}),x(t_{2})]}{2i}
\}e^{ik_{1}x_{1}}K(2),  \label{4.10}
\end{equation}
which is of the form (\ref{4.3}) with $n=2$. As a final step, we use the
Baker-Campbell-Hausdorf formula (\ref{B.3}) to write $
e^{ik_{1}x(t_{1})}e^{ik_{2}x(t_{2})}e^{k_{1}k_{2}[x(t_{1}),x(t_{2})]}=e^{i
\{k_{1}x(t_{1})+k_{2}x(t_{2})\}}$ and obtain the form (\ref{4.2}) with $n=2$.

For the general case, the argument goes in the same way. Assuming the form (
\ref{4.3}) for smaller $n$, we write the definition (\ref{4.1}) in the form
\begin{equation}
\xi (1,\cdots ,n)=K(2,\cdots ,n)\left\langle \int_{-\infty }^{\infty
}dx_{1}f^{\ast
}\{x_{1}-x(t_{1})\}e^{i\{k_{1}x_{1}+\sum_{l=2}^{n}k_{l}x(t_{l})\}}f
\{x_{1}-x(t_{1})\}\right\rangle .  \label{4.11}
\end{equation}
Then we bring the exponential factor to the right, using the theorem (\ref
{4.9}) as in Eq. (\ref{4.8}). Then, shifting the variable of integration and
the using the Baker-Campbell-Hausdorf formula in the exponential factor, we
get the form (\ref{4.3}).

\subsection{Example: Wave packet spreading}

Here we consider the case of two successive measurements, each with a
measurement function of the form (\ref{3.16}), corresponding to a Gaussian
slit. First, we consider a single measurement with
\begin{equation}
f(x_{1})=\frac{1}{(2\pi \sigma _{1}^{2})^{1/4}}\exp \{-\frac{x_{1}^{2}}{
4\sigma _{1}^{2}}\}.  \label{4.12}
\end{equation}
With this we use the standard Gaussian integral (\ref{B.1}), to evaluate the
integral expression (\ref{4.6}). We find
\begin{equation}
K(1)=\exp \{-\frac{1}{2}\sigma _{1}^{2}k_{1}^{2}\}.  \label{4.13}
\end{equation}
Using the Gaussian property (\ref{B.6}) to evaluate the expectation in (\ref
{4.5}), we find
\begin{equation}
\left\langle e^{ik_{1}x(t_{1})}\right\rangle =\exp \{-\frac{1}{2}
\left\langle x^{2}\right\rangle k_{1}^{2}\}.  \label{4.14}
\end{equation}
With this, we find
\begin{equation}
\xi (1)=\exp \{-\frac{1}{2}\sigma ^{2}k_{1}^{2}\},  \label{4.15}
\end{equation}
where we have introduced
\begin{equation}
\sigma ^{2}=\left\langle x^{2}\right\rangle +\sigma _{1}^{2}.  \label{4.16}
\end{equation}
Finally, we invert the definition (\ref{4.1}) of the characteristic function
to write
\begin{equation}
W(1)=\int_{-\infty }^{\infty }\frac{dk_{1}}{2\pi }\xi (1)e^{-ik_{1}x_{1}}.
\label{4.17}
\end{equation}
This again is a standard Gaussian integral and we obtain the result
\begin{equation}
W(1)=\frac{1}{\sqrt{2\pi \sigma ^{2}}}\exp \{-\frac{x_{1}^{2}}{2\sigma ^{2}}
\}.  \label{4.18}
\end{equation}
Thus the probability distribution associated with a single measurement is a
Gaussian whose variance is the sum of that of the instrument and that of the
underlying quantum state of the particle. We have presented the steps
leading to this result in detail since these are the steps that will be used
repeatedly in this and our later examples.

Consider now a pair of successive measurements, the first with measurement
function of the form (\ref{4.12}) the second of the same form but with the
index \textquotedblleft $1$\textquotedblright\ replaced by \textquotedblleft 
$2$\textquotedblright . Using the standard Gaussian integral to evaluate the
integral in the expression (\ref{4.10}) for $K(1,2)$, we find
\begin{equation}
K(1,2)=\exp \{-\frac{1}{2}\sigma _{1}^{2}k_{1}^{2}-\frac{1}{2}(\sigma
_{2}^{2}-\frac{[x(t_{1}),x(t_{2})]^{2}}{4\sigma _{1}^{2}})k_{2}^{2}\}.
\label{4.19}
\end{equation}
Then using the Gaussian property, we see that
\begin{equation}
\left\langle e^{i\{k_{1}x(t_{1})+k_{2}x(t_{2})\}}\right\rangle =\exp \{-
\frac{1}{2}\left\langle x^{2}\right\rangle
(k_{1}^{2}+k_{2}^{2})-c(t_{2}-t_{1})k_{1}k_{2}\}.  \label{4.20}
\end{equation}
Here we have introduced the correlation (\ref{A.9}). Putting these together,
using the expression (\ref{4.2}) for the characteristic function, we can
write
\begin{equation}
\xi (1,2)=\exp \{-\frac{1}{2}(\sigma ^{2}k_{1}^{2}+2\sigma \tau \rho
k_{1}k_{2}+\tau ^{2}k_{2}^{2}),  \label{4.21}
\end{equation}
where (note the misprint in the Eq. (7.18) of \cite{ford86})
\begin{eqnarray}
\sigma ^{2} &=&\left\langle x^{2}\right\rangle +\sigma _{1}^{2},  \notag \\
\tau ^{2} &=&\left\langle x^{2}\right\rangle +\sigma _{2}^{2}-\frac{
[x(t_{1}),x(t_{2})]^{2}}{4\sigma _{1}^{2}},  \notag \\
\sigma \tau \rho &=&c(t_{2}-t_{1}).  \label{4.22}
\end{eqnarray}
Note that $\sigma ^{2}$ is the same quantity (\ref{4.16}) that appears in
the single measurement function. The two measurement distribution function
is given by
\begin{equation}
W(1,2)=\int_{-\infty }^{\infty }\frac{dk_{1}}{2\pi }\int_{-\infty }^{\infty }
\frac{dk_{2}}{2\pi }\xi (1,2)e^{-i(k_{1}x_{1}+k_{2}x_{2})}.  \label{4.23}
\end{equation}
With the form (\ref{4.21}) of $\xi (1,2)$ we can perform the integration
using the multidimensional form (\ref{B.2}) of the standard Gaussian
integral. The result is
\begin{equation}
W(1,2)=\frac{1}{2\pi \sigma \tau \sqrt{1-\rho ^{2}}}\exp \{-\frac{\tau
^{2}x_{1}^{2}-2\sigma \tau \rho x_{1}x_{2}+\sigma ^{2}x_{2}^{2}}{2\sigma
^{2}\tau ^{2}(1-\rho ^{2})}\}.  \label{4.24}
\end{equation}

Here we remark first of all that the lack of symmetry of the quantum
distribution is obvious: $W(1,2)\neq W(2,1)$. The exception is when the
commutator vanishes. Note that the symmetry, or lack of it, is with respect
to interchange of the labels, that is, one must interchange not only $
x_{1}\leftrightarrows x_{2}$ and $t_{1}\leftrightarrows t_{2}$ but also $
\sigma _{1}\leftrightarrows \sigma _{2}$. Another aspect of this asymmetry
is that it is possible to make the last measurement one of perfect
precision, that is, put $\sigma _{2}=0$.

A second remark is that the time dependence is only through the time
difference $t_{2}-t_{1}$. This is a general feature, independent of the form
of the measurement function, It arises from the time-translation invariance
of the equilibrium state.

Finally, we remark that for widely separated times ($t_{2}-t_{1}\rightarrow
\infty $) the correlation and the commutator for the oscillator vanish. Then
we see that $W(1,2)\rightarrow W(1)W(2)$. This is a special case of the 
\emph{cluster property} of the quantum joint distribution functions, a
property they share with the classical functions. Whenever the time between
any two successive measurements is large, the quantum joint distribution
function factors into a product of distribution functions, the one
corresponding to the earlier times, the other to the later times. The
exception would be the case of a free particle, since in that case there is
no approach to an equilibrium value of $\left\langle x^{2}\right\rangle $.

\section{The probability distribution}

\label{sec:five}

In elementary quantum mechanics one interprets the absolute square of the
wave function as the probability distribution of the particle position. That
is, $\mathcal{P}(x,t)=\left\vert \psi (x,t)\right\vert ^{2}$, where $
\mathcal{P}dx$ is the probability of finding the particle in the interval $
dx $ at time $t$. In our discussion this probability distribution becomes
the conditional probability, specialized such that the second measurement is
a perfect measurement. That, is, the probability distribution is given by
\begin{equation}
\mathcal{P}(x_{2},t_{2}-t_{1})\equiv \left\{ \frac{W(1,2)}{W(1)}\right\}
_{\sigma _{2}=0}.  \label{5.1}
\end{equation}
Here we exhibit only the dependence on the final particle position and the
time difference. (Due to the time-translation invariance of the equilibrium
state only the time difference appears.) The picture we have is that the
initial state is prepared by the first measurement, a measurement made on
the equilibrium state, and that the second (perfect) measurement samples the
state. Of course, in the special case of a particle not interacting with the
bath and at temperature zero, this reduces to the elementary quantum
mechanics prescription.

If we use the expression (\ref{3.29}) for the joint probability
distributions, we see that we can write
\begin{equation}
\mathcal{P}(x_{2},t_{2}-t_{1})=\frac{\left\langle f(1)^{\dag }\delta
(x_{2}-x(t_{2}))f(1)\right\rangle }{\left\langle f(1)^{\dag
}f(1)\right\rangle }.  \label{5.2}
\end{equation}
While we can use the method described in Section \ref{sec:four} to calculate 
$W(1)$ and $W(1,2)$ and then put the results in the definition (\ref{5.1}),
it is just as well to evaluate the expression (\ref{5.2}) directly. For this
purpose we introduce the integral expression for the delta-function,
\begin{equation}
\delta (x)=\int_{-\infty }^{\infty }\frac{dP}{2\pi \hbar }e^{ixP/\hbar },
\label{5.3}
\end{equation}
with which we can write,
\begin{equation}
\mathcal{P}(x_{2},t_{2}-t_{1})=\int_{-\infty }^{\infty }\frac{dP}{2\pi \hbar 
}\frac{\left\langle f(1)^{\dag }e^{-ix(t_{2})P/\hbar }f(1)\right\rangle }{
\left\langle f(1)^{\dag }f(1)\right\rangle }e^{ix_{2}P/\hbar }.  \label{5.4}
\end{equation}
Now we use the formula (\ref{B.4}) to write
\begin{equation}
f(1)^{\dag }e^{-ix(t_{2})P/\hbar }f(1)=f^{\ast
}(x(t_{1})-x_{1})f(x(t_{1})-x_{1}-GP)e^{-ix(t_{2})P/\hbar },  \label{5.5}
\end{equation}
where we have used the relation (\ref{A.7}) to write the commutator in terms
of the green function, $G=G(t_{2}-t_{1})$. Next we again use the integral
expression (\ref{5.3}) for the delta-function to write
\begin{eqnarray}
f^{\ast }(x(t_{1})-x_{1})f(x(t_{1})-x_{1}-GP) &=&\int_{-\infty }^{\infty
}dx_{1}^{\prime }f^{\ast }(x_{1}^{\prime }+\frac{GP}{2})f(x_{1}^{\prime }-
\frac{GP}{2})  \notag \\
&&\times \int_{-\infty }^{\infty }\frac{dP^{\prime }}{2\pi \hbar }
e^{i(x_{1}+x_{1}^{\prime }-x(t_{1})+\frac{GP}{2})P^{\prime }/\hbar }.
\label{5.6}
\end{eqnarray}
Introducing this in (\ref{5.5}) and using the Baker-Campbell-Hausdorf
formula (\ref{B.3}) we can write
\begin{eqnarray}
\left\langle f(1)^{\dag }e^{-ix(t_{2})P/\hbar }f(1)\right\rangle
&=&\int_{-\infty }^{\infty }dx_{1}^{\prime }f^{\ast }(x_{1}^{\prime }+\frac{
GP}{2})f(x_{1}^{\prime }-\frac{GP}{2})  \notag \\
&&\times \int_{-\infty }^{\infty }\frac{dP^{\prime }}{2\pi \hbar }
e^{i(x_{1}+x_{1}^{\prime })P^{\prime }/\hbar }\left\langle
e^{-i(x(t_{1})P^{\prime }+x(t_{2})P)/\hbar }\right\rangle .  \label{5.7}
\end{eqnarray}
The Gaussian property (\ref{B.6}) allows us to write
\begin{equation}
\left\langle e^{-i(x(t_{1})P^{\prime }+x(t_{2}))P)/\hbar }\right\rangle
=\exp \{-\frac{\left\langle x^{2}\right\rangle (P^{\prime
2}+P^{2})+2cPP^{\prime }}{2\hbar ^{2}}\},  \label{5.8}
\end{equation}
where $c=c(t_{2}-t_{1})$ is the correlation (\ref{A.9}). Finally, with this
result, the integral over $P^{\prime }$ in (\ref{5.7}) is a standard
Gaussian integral (\ref{B.1}) and we obtain
\begin{eqnarray}
\left\langle f(1)^{\dag }e^{-ix(t_{2})P/\hbar }f(1)\right\rangle &=&\exp (-
\frac{(\left\langle x^{2}\right\rangle ^{2}-c^{2})P^{2}}{2\left\langle
x^{2}\right\rangle \hbar ^{2}}\}\int_{-\infty }^{\infty }dx_{1}^{\prime
}f^{\ast }(x_{1}^{\prime }+\frac{GP}{2})f(x_{1}^{\prime }-\frac{GP}{2}) 
\notag \\
&&\times \frac{\exp \{-\frac{(x_{1}+x_{1}^{\prime })^{2}}{2\left\langle
x^{2}\right\rangle }-i\frac{c}{\left\langle x^{2}\right\rangle }
(x_{1}+x_{1}^{\prime })\frac{P}{\hbar }\}}{\sqrt{2\pi \left\langle
x^{2}\right\rangle }}.  \label{5.9}
\end{eqnarray}
Here we should recall that $c=c(t_{2}-t_{1})$ is the correlation (\ref{A.9})
and $G=G(t_{2}-t_{1})$ is the Green function (\ref{A.5}). This is the key
result of this section, valid for any form of the measurement function $f$.
Dividing this result by its value for $P=0$ we get the integrand in the
expression (\ref{5.4}) for the probability distribution. We next consider
some examples corresponding to different choices of the measurement function.

\subsection{Example: Wave packet spreading}

Again, we consider wave packet spreading with an initial measurement
corresponding to a single Gaussian slit, the measurement function being that
given in Eq. (\ref{4.12}) \ There is no need to evaluate the general
expression (\ref{5.9}) since we already have the expressions (\ref{4.18}) for
$W(1)$ and (\ref{4.24}) for $W(1,2)$. Putting these in (\ref{5.1}) we can
write
\begin{equation}
\mathcal{P}(x,t)=\frac{1}{\sqrt{2\pi w^{2}(t)}}\exp \{-\frac{(x-\bar{x}
(t))^{2}}{2w^{2}(t)}\}.  \label{5.10}
\end{equation}
This is a Gaussian distribution with center $\bar{x}(t)$ and variance $
w^{2}(t)$, where
\begin{eqnarray}
\bar{x}(t) &=&\frac{\tau \rho }{\sigma }x_{1}  \notag \\
&=&\frac{c(t)}{\left\langle x^{2}\right\rangle +\sigma _{1}^{2}}x_{1}, 
\notag \\
w^{2}(t) &=&\tau ^{2}(1-\rho ^{2})  \notag \\
&=&\left\langle x^{2}\right\rangle -\frac{[x(0),x(t)]^{2}}{4\sigma _{1}^{2}}-
\frac{c^{2}(t)}{\left\langle x^{2}\right\rangle +\sigma _{1}^{2}}.
\label{5.11}
\end{eqnarray}
Here we we should again recall that $c(t)$ is the correlation (\ref{A.9}).

As a first consideration, we note that $\mathcal{P}(x,0)$ is the probability
distribution for the particle distribution immediately after the first
measurement. Since $c(0)=\left\langle x^{2}\right\rangle $ and the
commutator vanishes at $t=0$. we see that the center and variance of the
initial distribution are
\begin{eqnarray}
\bar{x}(0) &=&\frac{\left\langle x^{2}\right\rangle }{\left\langle
x^{2}\right\rangle +\sigma _{1}^{2}}x_{1},  \notag \\
w^{2}(0) &=&\frac{\left\langle x^{2}\right\rangle \sigma _{1}^{2}}{
\left\langle x^{2}\right\rangle +\sigma _{1}^{2}}.  \label{5.12}
\end{eqnarray}
As one can easily verify, this initial distribution corresponds to the
product of a Gaussian distribution of \ variance $\left\langle
x^{2}\right\rangle $ centered at the origin with one of variance $\sigma
_{1}^{2}$ centered at $x_{1}$. That is, the initial distribution corresponds
to the wave packet formed when the equilibrium state of the oscillator is
passed through a Gaussian slit of width $\sigma _{1}$ centered at $x_{1}$.

\subsubsection{Free particle}

The free particle coupled to the bath in the absence of the oscillator
potential corresponds to the limit $\left\langle x^{2}\right\rangle
\rightarrow \infty $. The point here is simple: the oscillator force can be
neglected near the center and the motion will be that of a free particle.
Noting that $c(t)=\left\langle x^{2}\right\rangle -s(t)/2$, where $s(t)$ is
the mean square displacement and remains finite in the limit, we find that
the center and variance (\ref{5.11}) of the probability distribution become
\begin{eqnarray}
\bar{x}(t) &=&x_{1}  \notag \\
w^{2}(t) &=&\sigma _{1}^{2}+s(t)-\frac{[x(0),x(t)]^{2}}{4\sigma _{1}^{2}}.
\label{5.13}
\end{eqnarray}
With the commutator expressed in terms of the Green function, this
expression for free particle wave packet spreading corresponds to that
obtained using path integral methods by Hakim and Ambegoakar \cite{hakim85}.
For a free particle not interacting with the bath and at temperature zero,
in which case $s(t)=0$ and $[x(0),x(t)]=i\hbar t/m$, this reduces to the
well known expression for wave packet spreading found in elementary quantum
textbooks.

\subsubsection{Displaced ground state distribution}

Another limit of interest is that in which $\sigma _{1}^{2}\rightarrow
\infty $, while at the same time $x_{1}\rightarrow \infty $ such that $
x_{0}=\left\langle x^{2}\right\rangle x_{1}/\sigma _{1}^{2}$ is fixed. In
this limit, the center and variance (\ref{5.11}) become 
\begin{eqnarray}
\bar{x}(t) &=&\frac{c(t)}{\left\langle x^{2}\right\rangle }x_{0},  \notag \\
w^{2}(t) &=&\left\langle x^{2}\right\rangle .  \label{5.14}
\end{eqnarray}
The probability distribution (\ref{5.12}) therefore is that of a displaced
equilibrium state. That is,
\begin{equation}
\mathcal{P}(x,t)=\mathcal{P}_{\text{eq}}(x-\bar{x}(t)),  \label{5.15}
\end{equation}
where 
\begin{equation}
\mathcal{P}_{\text{eq}}(x)=\frac{1}{\sqrt{2\pi \left\langle
x^{2}\right\rangle }}\exp \{-\frac{x^{2}}{2\left\langle x^{2}\right\rangle }
\}  \label{5.16}
\end{equation}
is the probability distribution of the equilibrium state. Thus, the wave
packet moves without spreading, the variance being \ that of the equilibrium
state of the oscillator. The center is initially at $x_{0}$ and
asymptotically approaches the origin as the equilibrium state is reached. We
shall come back to discuss this result further in Section \ref{sec:five}
when we discuss spreading of an initial coherent state.

\subsection{Example: \textquotedblleft Schr\"{o}dinger
cat\textquotedblright\ state}

\label{sec:five_B}

Here we consider the case where the initial measurement forms two separated
wave packets. The first measurement function then has the form
\begin{equation}
f(1)=\frac{\exp \{-\frac{(x_{1}-d/2)^{2}}{4\sigma _{1}^{2}}\}+\exp \{-\frac{
(x_{1}+d/2)^{2}}{4\sigma _{1}^{2}}\}}{[8\pi \sigma
_{1}^{2}(1+e^{-d^{2}/8\sigma _{1}^{2}})^{2}]^{1/4}}.  \label{5.17}
\end{equation}
With this form of the measurement function the integration in (\ref{5.9})
involves only the standard Gaussian integral (\ref{B.1}). Note that $x_{1}$
is the position of the center of the instrument, which should be chosen to
be zero if we wish the wave packet pair to be symmetrically placed about the
origin. We then find
\begin{eqnarray}
\left( \frac{\left\langle f(1)^{\dag }e^{i(x_{2}-x(t_{2}))P/\hbar
}f(1)\right\rangle }{\left\langle f(1)^{\dag }f(1)\right\rangle }\right)
_{x_{1}=0} &=&\frac{\exp \{-\frac{w^{2}P^{2}}{2\hbar ^{2}}+ix_{2}\frac{P}{
\hbar }\}}{1+\exp \{-\frac{\left\langle x^{2}\right\rangle d^{2}}{8\sigma
_{1}^{2}(\left\langle x^{2}\right\rangle +\sigma _{1}^{2})}\}}  \notag \\
&&\times (\cos \frac{P\bar{d}}{2\hbar }+\exp \{-\frac{\left\langle
x^{2}\right\rangle d^{2}}{8\sigma _{1}^{2}(\left\langle x^{2}\right\rangle
+\sigma _{1}^{2})}\}\cosh \frac{GPd}{4\sigma _{1}^{2}}),  \label{5.18}
\end{eqnarray}
where $w^{2}=w^{2}(t_{2}-t_{1})$ is given in (\ref{5.11}) and $\bar{d}=\bar{d
}(t_{2}-t_{1})$ with
\begin{equation}
\bar{d}(t)=\frac{c(t)}{\left\langle x^{2}\right\rangle +\sigma _{1}^{2}}d.
\label{5.19}
\end{equation}
While there is no difficulty evaluating the integral expression (\ref{5.4})
with this expression for the integrand, our interest will be in the limits
of a free particle or, for the oscillator, a displaced ground state pair, in
which case it is simpler to first evaluate the limits of the above
expression and then evaluate the integral.

\subsubsection{Free particle}

As in the above example of wave packet spreading, we obtain the case of a
free particle coupled to the bath in the absence of the oscillator potential
by forming the limit $\left\langle x^{2}\right\rangle \rightarrow \infty $.
Forming this limit of the expression (\ref{5.18}) then putting the result in
the expression (\ref{5.4}) for the probability distribution and performing
the integral with the standard Gaussian formula (\ref{B.1}) we find
\begin{eqnarray}
\mathcal{P}(x,t) &=&\frac{1}{2(1+e^{-d^{2}/8\sigma _{1}^{2}})}\left\{ \frac{
\exp \{-\frac{(x-d/2)^{2}}{2w^{2}}\}}{\sqrt{2\pi w^{2}}}+\frac{\exp \{-\frac{
(x+d/2)^{2}}{w^{2}}\}}{\sqrt{2\pi w^{2}}}\right.  \notag \\
&&\left. +2a\frac{\exp \{-\frac{x^{2}+d^{2}/4}{2w^{2}}\}}{\sqrt{2\pi w^{2}}}
\cos \frac{[x(0),x(t)]xd}{4i\sigma _{1}^{2}w^{2}}\right\} ,  \label{5.20}
\end{eqnarray}
where now $w^{2}(t)$ is given by the free particle form (\ref{5.13}) and we
have used the relation (\ref{A.7}) to reintroduce the commutator. In this
expression $a(t)$ is the attenuation coefficient, given by
\begin{equation}
a(t)=\exp \{-\frac{s(t)d^{2}}{8\sigma _{1}^{2}w^{2}(t)}\}.  \label{5.21}
\end{equation}
This expression for the probability distribution is the same as that derived
in an earlier brief communication \cite{ford01}.

This probability distribution is the sum of three contributions,
corresponding to the three terms within the braces. The first two are
probability distributions of the form (\ref{5.12}) corresponding to a pair
of single slits positioned at $\pm d/2$, while the third term (that
involving the cosine) is an interference term. The attenuation coefficient
is a measure of the size of the interference term and is defined as the
ratio of the amplitude of the interference term to twice the geometric mean
of the other two terms. The point here is perhaps best seen if we look first
at the case of a particle\ without dissipation and at zero temperature. Then 
$s(t)=0$ and $[x(0),x(t)]=i\hbar t/m$. The resulting probability
distributions are shown in Figure \ref{qdist_fig1}. There $\mathcal{P}(x,0)$
is the initial probability distribution, which depends only on the initial
measurement function and is therefore the same whether or not there is
dissipation. In this initial distribution the interference term corresponds
to a miniscule peak at the origin, so small that it does not show in the
plot. In the same figure $\mathcal{P}(x,t)$ is the distribution at a time $t$
such that the width of the individual wave packets has increased by a factor
of roughly $3$, while $\mathcal{P}_{0}(x,t)$ is the same distribution but
with the attenuation factor $a(t)$ set equal to zero. We emphasize that at
this later time the amplitude of the interference term is of the order of
that of the other terms, despite the fact that initially it is negligibly
small. The difference between $\mathcal{P}(x,t)$, where the attenuation
factor is unity and the interference term is present, and $\mathcal{P}
_{0}(x,t)$, where the interference term is absent, is what is called
decoherence. Thus, the attenuation coefficient corresponds to the
traditional measure of coherence in terms of the relative amplitude of an
observed interference term.\cite{zernike}

In the presence of dissipation the attenuation factor decays rapidly when
the separation $d$ of the wave packets is large. Here by \textquotedblleft
large\textquotedblright\ we mean not only large compared with the slit width 
$\sigma _{1}$ but also large compared with the mean de Broglie wavelength $
\bar{\lambda}=\hbar /m\sqrt{\left\langle \dot{x}^{2}\right\rangle }$. To
obtain the short time behavior in this case, we note that, as we have seen
in Section, \ref{sec:three}, for short times the mean square displacement $
s(t)\cong \left\langle \dot{x}^{2}\right\rangle t^{2}$. Then, since for
short times $w^{2}(t)\cong \sigma _{1}^{2}$, we see that for short times,
\begin{equation}
a(t)\cong e^{-t^{2}/2\tau _{\text{d}}^{2}},  \label{5.22}
\end{equation}
where $\tau _{\text{d}}$, the decoherence. time, is given by
\begin{equation}
\tau _{\text{d}}=\frac{2\sigma _{1}^{2}}{\sqrt{\left\langle \dot{x}
^{2}\right\rangle }d}.  \label{5.23}
\end{equation}
In the high temperature case, where $\left\langle \dot{x}^{2}\right\rangle
=kT/m$, this is the result for the decoherence. time obtained previously 
\cite{ford01}, but the result holds equally well at zero temperature, where $
\left\langle \dot{x}^{2}\right\rangle $ is given in Eq. (\ref{A.29}). In
either case, the decoherence. time is very short when the separation $d$ of
the pair of wave packets is large.

\subsubsection{Displaced ground state pair}

As in the above example of wave packet spreading, we obtain a relatively
simple expression for the oscillator case in the limit $\sigma
_{1}^{2}\rightarrow \infty $ and $d\rightarrow \infty $ such that $
d_{0}=\left\langle x^{2}\right\rangle d/\sigma _{1}^{2})$ is fixed. Forming
this limit of the result (\ref{5.18}) and then evaluating the expression (
\ref{5.4}) for the probability distribution, we find
\begin{eqnarray}
\mathcal{P}(x,t) &=&\frac{1}{2(1+e^{-d_{0}^{2}/8\left\langle
x^{2}\right\rangle })}\left\{ \mathcal{P}_{\text{eq}}(x-\frac{\bar{d}}{2})+
\mathcal{P}_{\text{eq}}(x+\frac{\bar{d}}{2})\right.  \notag \\
&&\left. 2a(t)e^{-\bar{d}^{2}(t)/8\left\langle x^{2}\right\rangle }\mathcal{P
}_{\text{eq}}(x)\cos \frac{[x(0),x(t)]xd_{0}}{4i\left\langle
x^{2}\right\rangle ^{2}}\right\} ,  \label{5.24}
\end{eqnarray}
where $\mathcal{P}_{\text{eq}}$ is the equilibrium distribution (\ref{5.16})
and now
\begin{equation}
\bar{d}(t)=\frac{c(t)}{\left\langle x^{2}\right\rangle }d_{0},  \label{5.25}
\end{equation}
while here the attenuation coefficient is given by
\begin{equation}
a(t)=\exp \{-\frac{d_{0}^{2}}{8\left\langle x^{2}\right\rangle }(1-\frac{
c^{2}}{\left\langle x^{2}\right\rangle ^{2}}+\frac{[x(0),x(t)]^{2}}{
4\left\langle x^{2}\right\rangle ^{2}})\}.  \label{5.26}
\end{equation}

In connection with this result we remark first that initially the
distribution is of the same form as in the free particle case, with two
Gaussian peaks at $\pm d_{0}/2$ and a miniscule central peak. The difference
is in the motion: the two peaks drift back and forth against each other
without spreading, eventually arriving at the origin. The interference term
therefore has a different effect (sometimes called a \textquotedblleft
quantum carpet\textquotedblright\ \cite{kaplan98}) but we have nevertheless
introduced the attenuation coefficient in the same way as the ratio of the
coefficient of the cosine to twice the geometric mean of the first two
terms. For very short times, $c(t)\cong \left\langle x^{2}\right\rangle -
\frac{1}{2}\left\langle \dot{x}^{2}\right\rangle t^{2}$ and $
[x(0),x(t)]\cong i\hbar t/m$, so $a(t)$ is of the same form (\ref{5.22}) but
now with
\begin{equation}
\tau _{\text{d}}=\frac{2\left\langle x^{2}\right\rangle }{\sqrt{\left\langle 
\dot{x}^{2}\right\rangle -\hbar ^{2}/4m^{2}\left\langle x^{2}\right\rangle }
d_{0}}.  \label{5.27}
\end{equation}
Note that the uncertainty principle tells us that $m^{2}\left\langle \dot{x}
^{2}\right\rangle \left\langle x^{2}\right\rangle \geq \hbar ^{2}/4$, so the
argument of the square root is always positive..

\section{The Wigner function}

\label{sec:six}

The Wigner function is the analog of the probability distribution in which
the second measurement is a perfect measurement of position \emph{and}
momentum. Of course, a quantum particle cannot have simultaneously a precise
position and momentum, so this last cannot be a proper quantum measurement
of the general form (\ref{3.7}). The Wigner function is therefore not a
probability distribution but rather what is called a \textquotedblleft
quasiprobability distribution function\textquotedblright . To get the Wigner
function corresponding to the density matrix at time $t_{2}$, which we
denote by $\mathcal{W}(q,p;t_{2}-t_{1})$, we make the replacement $f^{\dag
}(2)f(2)\rightarrow ``\delta (q-x(t_{2})\delta (p-m\dot{x}(t_{2}))^{\prime
\prime }=F(q-x(t_{2}),p-m\dot{x}(t_{2}))$, where
\begin{equation}
F(q,p)=\frac{1}{(2\pi \hbar )^{2}}\int_{-\infty }^{\infty }dQ\int_{-\infty
}^{\infty }dPe^{i(Pq+Qp)/\hbar }.  \label{6.1}
\end{equation}
That is, in place of the general formula (\ref{5.2}) for the probability
distribution we have its generalization to quantum phase space given by the
general formula 
\begin{eqnarray}
\mathcal{W}(q,p;t_{2}-t_{1}) &=&\frac{\left\langle f^{\dag
}(1)F(q-x(t_{2}),p-m\dot{x}(t_{2}))f(1)\right\rangle }{\left\langle f^{\dag
}(1)f(1)\right\rangle }  \notag \\
&=&\int_{-\infty }^{\infty }\frac{dQ}{2\pi \hbar }\int_{-\infty }^{\infty }
\frac{dP}{2\pi \hbar }e^{i(Pq+Qp)/\hbar }\frac{\left\langle f^{\dag
}(1)e^{-i(x(t_{2})P+m\dot{x}(t_{2})Q)/\hbar }f(1)\right\rangle }{
\left\langle f^{\dag }(1)f(1)\right\rangle }.  \label{6.2}
\end{eqnarray}

Note first of all that as a classical function $F(q,p)=\delta (q)\delta (p)$
. This expression would be unsatisfactory for our purposes, since in the
general formula (\ref{6.2}) the arguments of the delta functions would not
commute. The integral expression (\ref{6.1}) corresponds to the Fourier-von
Neumann representation of the classical operator. Second, we note that the
momentum operator is interpreted as the \emph{mechanical} momentum $m\dot{x}$
. This, as we have noted above, is in accord with the macroscopic
description of a dissipative system. The point here is that the canonical
momentum is an operator of the microscopic description, which for the same
macroscopic description may or may not be equal to the mechanical momentum 
\cite{ford88b}. Finally, we should emphasize that the Wigner function is not
a probability distribution as discussed in Sec. \ref{sec:three}. For this
reason we use a calligraphic $\mathcal{W}$ to help keep this in mind. This
formula for the Wigner function is unique in the sense that it satisfies
certain general requirements such as that it be a real function, that the
integral over $p$ or $q$ must give the corresponding probability
distribution in position or momentum, etc. For a thorough discussion of the
Wigner function as it has appeared in the literature, see the review article
of Hillery et al. \cite{hillery84} (see especially their Eq. (2.45).

It is convenient to introduce the Fourier transform of the Wigner function, 
\begin{equation}
\mathcal{\tilde{W}}(Q,P;t)=\int_{-\infty }^{\infty }dq\int_{-\infty
}^{\infty }dpe^{-i(Pq+Qp)/\hbar }\mathcal{W}(q,p;t).  \label{6.3}
\end{equation}
This Fourier transform is what in the literature is called the
\textquotedblleft characteristic function\textquotedblright\ \cite{hillery84}
. We adopt this convenient terminology, but warn that this Wigner
characteristic function should not be confused with the quantum analog of
the characteristic functions of classical probability introduced in Section 
\ref{sec:four}. The inverse Fourier transform is 
\begin{equation}
\mathcal{W}(q,p;t)=\int_{-\infty }^{\infty }\frac{dQ}{2\pi \hbar }
\int_{-\infty }^{\infty }\frac{dP}{2\pi \hbar }\mathcal{\tilde{W}}
(Q,P;t)e^{i(Pq+Qp)/\hbar }.  \label{6.4}
\end{equation}
Comparing this with the general formula (\ref{6.2}), we obtain by inspection
a simple formula for the Wigner characteristic function:
\begin{equation}
\mathcal{\tilde{W}}(Q,P;t_{2}-t_{1})=\frac{\left\langle f^{\dag
}(1)e^{-i(x(t_{2})P+m\dot{x}(t_{2})Q)/\hbar }f(1)\right\rangle }{
\left\langle f^{\dag }(1)f(1)\right\rangle }.  \label{6.5}
\end{equation}
This is the key result of this section, valid for any form of the
measurement function $f$. As we next shall show, it allows us to readily
calculate the Wigner function for a variety of examples.

\subsection{Example: Equilibrium Wigner function}

As a first simple example we consider the equilibrium Wigner function, which
we denote by $\mathcal{W}_{\text{eq}}(q,p)$, and which we get when we make
no initial measurement. This corresponds to $f(1)\rightarrow 1$, and for
this case the formula (\ref{6.5}) for the Wigner characteristic function
becomes
\begin{equation}
\mathcal{\tilde{W}}_{\text{eq}}(Q,P)=\left\langle e^{-i(x(t_{2})P+m\dot{x}
(t_{2})Q)/\hbar }\right\rangle .  \label{6.6}
\end{equation}
Using the Gaussian formula (\ref{B.6}) we find 
\begin{equation}
\mathcal{\tilde{W}}_{\text{eq}}(Q,P)=\exp \{-\frac{1}{2\hbar ^{2}}
(\left\langle x^{2}\right\rangle P^{2}+m^{2}\left\langle \dot{x}
^{2}\right\rangle Q^{2})\},  \label{6.7}
\end{equation}
where we have used the fact that $\left\langle x\dot{x}+\dot{x}
x\right\rangle =0$. With this in (\ref{6.4}), the integrals are standard
Gaussian integrals and we find for the equilibrium Wigner function,
\begin{equation}
\mathcal{W}_{\text{eq}}(q,p)=\frac{1}{2\pi m\sqrt{\left\langle
x^{2}\right\rangle \left\langle \dot{x}^{2}\right\rangle }}\exp \{-\frac{
q^{2}}{2\left\langle x^{2}\right\rangle }-\frac{p^{2}}{2m^{2}\left\langle 
\dot{x}^{2}\right\rangle }\}.  \label{6.8}
\end{equation}
Here $\left\langle x^{2}\right\rangle $ and $\left\langle \dot{x}
^{2}\right\rangle $ are given in the expressions (\ref{A.30}) and (\ref{A.29}
). The familiar weak coupling form, well known as the equilibrium solution
of the master equation, results if we recall the relations for weak coupling
given in Eq. (\ref{A.39}).

\subsection{Example: Motion of a coherent state}

Coherent states are generally defined for the free oscillator by operating
on the oscillator ground state with the general displacement operator $\exp
\{i(mv_{0}x-x_{0}p)/\hbar \}$. The resulting coherent state corresponds to a
displaced ground state, centered at $x_{0}$ and moving with velocity $v_{0}$
. Here we define a generalized coherent state for an oscillator interacting
with a linear passive heat bath. The corresponding density matrix is
obtained by acting on the equilibrium density matrix with the measurement
function
\begin{equation}
f(1)=\exp \{im(v_{0}x(t_{1})-x_{0}\dot{x}(t_{1}))/\hbar \}.  \label{6.9}
\end{equation}
With this the expression (\ref{6.5}) for the Wigner characteristic function
becomes
\begin{equation}
\mathcal{\tilde{W}}(Q,P;t_{2}-t_{1})=\left\langle e^{-im(v_{0}x(t_{1})-x_{0}
\dot{x}(t_{1})/\hbar }e^{-i(x(t_{2})P+m\dot{x}(t_{2})Q)/\hbar
}e^{im(v_{0}x(t_{1})-x_{0}\dot{x}(t_{1}))/\hbar }\right\rangle .
\label{6.10}
\end{equation}
Use the Baker-Campbell-Hausdorf formula (\ref{B.3}) to write
\begin{eqnarray}
&&e^{-im(v_{0}x(t_{1})-x_{0}\dot{x}(t_{1}))/\hbar }e^{-i(x(t_{2})P+m\dot{x}
(t_{2})Q)/\hbar }e^{im(v_{0}x(t_{1})-x_{0}\dot{x}(t_{1}))/\hbar }  \notag \\
&=&e^{-i(x(t_{2})P+m\dot{x}(t_{2})Q)/\hbar }e^{m[x(t_{2})P+m\dot{x}
(t_{2})Q,v_{0}x(t_{1})-x_{0}\dot{x}(t_{1})]/\hbar ^{2}}.  \label{6.11}
\end{eqnarray}
With this, we find
\begin{equation}
\mathcal{\tilde{W}}(Q,P;t)=\exp \{-i\frac{1}{\hbar }\left( \bar{x}(t)P+m\bar{
v}(t)Q\right) \}\mathcal{\tilde{W}}_{\text{eq}}(Q,P),  \label{6.12}
\end{equation}
where, using Eq. (\ref{A.7}) to express the commutator in terms of the Green
function,
\begin{eqnarray}
\bar{x}(t) &=&m\dot{G}(t)x_{0}+mG(t)v_{0},  \notag \\
\bar{v}(t) &=&\frac{d\bar{x}(t)}{dt}=m\ddot{G}(t)x_{0}+m\dot{G}(t)v_{0}.
\label{6.13}
\end{eqnarray}
The Wigner function is given by the inverse Fourier transform (\ref{6.4}).
We find
\begin{equation}
\mathcal{W}(q,p;t)=\mathcal{W}_{\text{eq}}(q-\bar{x}(t),p-m\bar{v}(t)).
\label{6.14}
\end{equation}
Thus the Wigner function corresponding to an initial coherent state has the
form of an equilibrium Wigner function whose center moves according to the
equations (\ref{6.13}).

We should point out that the motion (\ref{6.13}) of this center is not the
solution of the mean of the quantum Langevin equation (\ref{A.1}). Rather,
it is the solution of the mean of the initial value Langevin equation \cite
{ford87},
\begin{equation}
m\frac{d^{2}\bar{x}}{dt^{2}}+\int_{0}^{t}dt^{\prime }\mu (t-t^{\prime })
\frac{d\bar{x}(t^{\prime })}{dt^{\prime }}+K\bar{x}=-\mu (t)\bar{x}(0),
\label{6.15}
\end{equation}
with initial data $\bar{x}(0)=x_{0}$ and $\bar{v}(t)=v_{0}$. The effect of
the term on the right hand side can be seen clearly in the Ohmic case, where
the Green function has the form (\ref{A.20}). With that form we see that $
\bar{x}(t+0^{+})=x_{0}$, while $\bar{v}(t+0^{+})=v_{0}-\gamma x_{0}$. Thus
the center of initial distribution in the $qp$ plane makes a jump in the $p$
direction, down (up) if $x_{0}$ is positive (negative), after which the
motion of the center is that of a damped harmonic oscillator. At all times
the shape of the distribution is that of a displaced thermal equilibrium
state of the oscillator.

The probability distribution is obtained by integrating over $p$,
\begin{equation}
\mathcal{P}(x,t)=\int_{-\infty }^{\infty }dp\mathcal{W}(x,p;t).  \label{6.16}
\end{equation}
With the expression (\ref{6.14}) for the Wigner function in which $\mathcal{W
}_{\text{eq}}$ is of the form (\ref{6.8}) this becomes
\begin{equation}
\mathcal{P}(x,;t)=\mathcal{P}_{\text{eq}}(x-\bar{x}(t)),  \label{6.17}
\end{equation}
where $\mathcal{P}_{\text{eq}}$ is the equilibrium distribution (\ref{5.16}
). This is exactly the form we encountered above in the example of wave
packet spreading, where the parameters for the displaced ground state are
given in Eq. (\ref{5.14}). In either case the probability distribution is
that of a displaced ground state. The difference lies in the motion of the
center, which is temperature independent in the case of the coherent state
but has a temperature dependent form for the displaced ground state. In
either case, of course, the distribution approaches that of equilibrium for
long times. The lesson we learn here is that the time dependence of the
approach to equilibrium depends on how the initial state is formed. In
Figure \ref{qdist_fig2} we plot $\bar{x}(t)$ for the two cases, the
displaced ground state motion being calculated at zero temperature. The
parameters chosen were $\gamma /\omega _{0}=10/13$ and $\Omega /\omega
_{0}=5 $, but despite this rather strong coupling, there is not much
difference between the two curves.

\subsection{Example: Coherent state pair}

\label{sec:six-C}

The idea here is to form an initial state like the \textquotedblleft Schr
\"{o}dinger cat\textquotedblright\ state discussed in Sec.\ref{sec:five_B}.
There the initial state was prepared with a pair of Gaussian slits. Here we
consider instead an initial state which is a superposition of two separated
coherent states. This is accomplished with a measurement function of the
simple form 
\begin{equation}
f(1)=\cos \frac{md\dot{x}(t_{1})}{2\hbar },  \label{6.18}
\end{equation}
which results in a superposed pair of generalized coherent states, centered
at $x=\pm d/2$ and each with zero velocity. The expression (\ref{6.5}) for
the Fourier transform of the normalized Wigner function at time $t_{2}$
therefore becomes 
\begin{equation}
\mathcal{\tilde{W}}(Q,P;t_{2}-t_{1})=\frac{\left\langle \cos \frac{md\dot{x}
(t_{1})}{2\hbar }e^{-i(x(t_{2})P+m\dot{x}(t_{2})Q)/\hbar }\cos \frac{md\dot{x
}(t_{1})}{2\hbar }\right\rangle }{\left\langle \cos ^{2}\frac{md\dot{x}
(t_{1})}{2\hbar }\right\rangle }.  \label{6.19}
\end{equation}
Now, using again the Baker-Campbell-Hausdorf formula (\ref{B.3}), we see
that 
\begin{eqnarray}
&&\cos \frac{md\dot{x}(t_{1})}{2\hbar }e^{-i(x(t_{2})P+m\dot{x}
(t_{2})Q)/\hbar }\cos \frac{md\dot{x}(t_{1})}{2\hbar }  \notag \\
&=&\frac{1}{4}(2e^{-i(x(t_{2})P+m\dot{x}(t_{2})Q)/\hbar }\cos \frac{md(\dot{G
}P+m\ddot{G}Q)}{2\hbar }  \notag \\
&&+e^{-i(x(t_{2})P+m\dot{x}(t_{2})Q-md\dot{x}(t_{1}))/\hbar
}+e^{-i(x(t_{2})P+m\dot{x}(t_{2})Q+md\dot{x}(t_{1}))/\hbar }),  \label{6.20}
\end{eqnarray}
where, to shorten the expression, we have used the expression (\ref{A.7}) to
express the commutator in terms of the Green function. Putting this in (\ref
{6.19}) and using the Gaussian property (\ref{B.6}) we find 
\begin{eqnarray}
\mathcal{\tilde{W}}(Q,P;t) &=&\frac{\exp \{-(\left\langle x^{2}\right\rangle
P^{2}+m^{2}\left\langle \dot{x}^{2}\right\rangle Q^{2})/2\hbar ^{2}\}}{
1+\exp \{-m^{2}\left\langle \dot{x}^{2}\right\rangle d^{2}/2\hbar ^{2}}
\left( \cos \frac{md(\dot{G}P+m\ddot{G}Q)}{2\hbar }\right.  \notag \\
&&\left. +\exp \{-\frac{m^{2}d^{2}\left\langle \dot{x}^{2}\right\rangle }{
2\hbar ^{2}}\}\cosh \frac{md(\dot{s}P+m\ddot{s}Q)}{2\hbar ^{2}}\right) ,
\label{6.21}
\end{eqnarray}
where $G=G(t)$ is the Green function (\ref{A.5}) and $s=s(t)$ is the mean
square displacement (\ref{A.8}). Putting this in (\ref{6.4}), the integral
is a two dimensional standard Gaussian (\ref{B.2}) and we find 
\begin{eqnarray}
\mathcal{W}(q,p;t) &=&\frac{1}{2(1+\exp \{-m^{2}\left\langle \dot{x}
^{2}\right\rangle d^{2}/2\hbar ^{2}\})}\left\{ \mathcal{W}_{\text{eq}}(q-
\frac{m\dot{G}d}{2},p-\frac{m^{2}\ddot{G}d}{2})\right.  \notag \\
&&+\mathcal{W}_{\text{eq}}(q+\frac{m\dot{G}d}{2},p+\frac{m^{2}\ddot{G}d}{2})
\notag \\
&&\left. +2e^{-A(t)}\mathcal{W}_{\text{eq}}(q,p)\cos \Phi (q,p;t)\right\} ,
\label{6.22}
\end{eqnarray}
where $\mathcal{W}_{\text{eq}}(q,p)$ is the equilibrium Wigner function (\ref
{6.8}) and we have introduced
\begin{eqnarray}
A(t) &=&\frac{m^{2}d^{2}\left\langle \dot{x}^{2}\right\rangle }{2\hbar ^{2}}
\left( 1-\frac{\dot{s}^{2}(t)}{4\left\langle x^{2}\right\rangle \left\langle 
\dot{x}^{2}\right\rangle }-\frac{\ddot{s}^{2}(t)}{4\left\langle \dot{x}
^{2}\right\rangle ^{2}}\right)  \notag \\
\Phi (q,p;t) &=&\left( \frac{m\dot{s}q}{\left\langle x^{2}\right\rangle }+
\frac{\ddot{s}p}{\left\langle \dot{x}^{2}\right\rangle }\right) \frac{d}{
2\hbar }.  \label{6.23}
\end{eqnarray}
This expression for the Wigner function is identical with that obtained by
Romero and Paz using path integral methods \cite{paz97}.

Viewed in the $qp$ plane, the expression (\ref{6.22}) for the Wigner
function shows three peaks: an outlying pair in the form of single coherent
states, centered initially at $q=\frac{d}{2}$, $p=0$, and an interference
peak, centered at the origin and modulated by the factor cos$\Phi $.
Initially, since $A(0)=0$, the amplitude of the interference peak is twice
that of either of the two outlying peaks. In general the interest is in the
case of a widely separated coherent state pair, that is, $d\gg \sqrt{
\left\langle x^{2}\right\rangle }$. In that case for very short times $A(t)$
becomes large and the interference peak is practically zero. This
disappearance of the interference peak is the phenomenon of decoherence. as
seen with the Wigner function. To be more explicit, with the expansion (\ref
{A.26}) in the expression (\ref{A.24}) for $s(t)$, we can readily evaluate
this expression for short times. In the limit of small bath relaxation time
this takes the simple form:
\begin{equation}
A(t)\cong -\frac{t^{2}}{2\tau _{\text{d}}^{2}}\log \frac{t}{\tau },
\label{6.24}
\end{equation}
where
\begin{equation}
\tau _{\text{d}}=\tau \sqrt{\frac{\pi \hbar }{\zeta d^{2}}}.  \label{6.25}
\end{equation}
Here $\zeta $ is the friction constant and $\tau $ is the bath relaxation
time, the parameters in the single relaxation time model. We can interpret $
\tau _{\text{d}}$, the time of the order of which the central peak in the
Wigner distribution vanishes, as a decoherence. time. Note that this
expression for the decoherence. time is very different from that for the
displaced ground state pair given in Eq. (\ref{5.27}), despite the
similarity of the initial states. The qualitative nature of the phenomenon
is the same: the rapid disappearance of an interference term.

Since the Wigner function is not directly observable, we should consider the
probability distribution. That is, we put the expression (\ref{6.22}) for
the Wigner function in the integral (\ref{6.16}) for the probability
distribution to obtain
\begin{eqnarray}
\mathcal{P}(x,t) &=&\frac{1}{2(1+\exp \{-m^{2}\left\langle \dot{x}
^{2}\right\rangle d^{2}/2\hbar ^{2}\})}\left( \mathcal{P}_{\text{eq}}(x-m
\dot{G}\frac{d}{2})+\mathcal{P}_{\text{eq}}(x+m\dot{G}\frac{d}{2})\right. 
\notag \\
&&\left. 2a\exp \left\{ -\frac{m^{2}\dot{G}^{2}d^{2}}{8\left\langle
x^{2}\right\rangle }\right\} \mathcal{P}_{\text{eq}}(x)\cos \frac{m\dot{s}dx
}{2\hbar \left\langle x^{2}\right\rangle }\right) ,  \label{6.26}
\end{eqnarray}
where $a=a(t)$ is the attenuation coefficient, now given by
\begin{equation}
a(t)=\exp \left\{ -\left( \frac{4m^{2}\left\langle \dot{x}^{2}\right\rangle
\left\langle x^{2}\right\rangle }{\hbar ^{2}}-\frac{m^{2}\dot{s}^{2}(t)}{
\hbar ^{2}}-m^{2}\dot{G}^{2}\right) \frac{d^{2}}{8\left\langle
x^{2}\right\rangle }\right\} .  \label{6.27}
\end{equation}
As in our discussion of the Schr\"{o}dinger cat state in Sec. \ref{sec:five}
, the attenuation coefficient is defined as the ratio of the coefficient of
the cosine term to the geometric mean of the first two terms and corresponds
to the traditional measure of coherence. Its disappearance is the phenomenon
of decoherence. as seen in the probability distribution. But here the
initial value of the attenuation coefficient,
\begin{equation}
a(0)=\exp \left\{ -\left( \frac{4m^{2}\left\langle \dot{x}^{2}\right\rangle
\left\langle x^{2}\right\rangle }{\hbar ^{2}}-1\right) \frac{d^{2}}{
8\left\langle x^{2}\right\rangle }\right\} ,  \label{6.28}
\end{equation}
is already vanishingly small for large separations (the uncertainty
principle tells us that the factor in the exponent is necessarily positive).
Indeed, for the Ohmic model, the mean square velocity is logarithmically
divergent, so the attenuation coefficient is identically zero for all times.
In any event, we would say that as seen in the probability distribution, the
decoherence. for a coherent state pair occurs initially and there is no
notion of a decoherence. time.

The earliest discussion of this problem of a displaced pair of coherent
states at zero temperature was that of Walls and Milburn,\cite{walls1985}
who based their discussion on the master equation. As we note in the last
paragraph of appendix \ref{appendix:A}, this would correspond to the
Weisskpof-Wigner approximation. In this approximation
\begin{equation}
A(t)=\frac{m\omega _{0}d^{2}}{4\hbar }\left( 1-e^{-\gamma t}\right)
\label{6.29}
\end{equation}
and
\begin{equation}
a(t)=\exp \{-A(t)\}.  \label{6.30}
\end{equation}
Note first of all that in this Weisskopf-Wigner approximation the decay of
coherence is identical whether viewed in the Wigner function or in the
probability distribution, \ However, the short time behavior \ is very
different from that of the exact result. As an illustration, in Figure \ref
{qdist_fig3} we compare the short time behavior of the exact result with
that of the Weisskopf-Wigner approximation as well as that of the weak
coupling approximation. In this range the weak coupling expression for $A(t)$
is just twice that from the Weisskopf-Wigner approximation, while both are
much larger than the exact result and would give a correspondingly much
shorter estimate of the decoherence. time. The conclusion to be drawn is
that the master equation can give misleading results at short times.

\subsection{Example: Squeezed state}

The squeezed state that appears in the quantum optics literature is obtained
by operating on the ground state of the free oscillator with the squeeze
operator \cite{scully97},
\begin{equation}
S=\exp \{\frac{r}{2}(e^{-i\theta }a^{2}-e^{i\theta }a^{\dag 2})\},
\label{6.31}
\end{equation}
where $r$ and $\theta $ are real parameters and $a=(m\omega _{0}x+ip)/\sqrt{
2m\hbar \omega _{0}}$ is the usual annihilation operator for the free
oscillator. Here we consider the so called ideal squeeze operator,
corresponding to $\theta =0$, and again replace the canonical momentum with
the mechanical momentum. The ideal squeeze operation would therefore
correspond to an initial measurement operator of the form 
\begin{equation}
f(1)=\exp \{i\frac{mr}{2\hbar }(x(t_{1})\dot{x}(t_{1})+\dot{x}
(t_{1})x(t_{1}))\}.  \label{6.32}
\end{equation}
Since this is a unitary operator, we see that 
\begin{equation}
\left\langle f^{\dag }(1)e^{-i(x(t_{2})P+m\dot{x}(t_{2})Q)/\hbar
}f(1)\right\rangle =\left\langle e^{-i(X(r,t_{2})P+m\dot{X}(r;t_{2})Q)/\hbar
}\right\rangle ,  \label{6.33}
\end{equation}
where we have introduced the operator 
\begin{equation}
X(r;t_{2})=f^{\dag }(1)x(t_{2})f(1).  \label{6.34}
\end{equation}

To evaluate this operator, form the derivative with respect to $r$, 
\begin{eqnarray}
\frac{\partial X(r;t_{2})}{\partial r} &=&f^{\dag }(1)\frac{im}{2\hbar }
[x(t_{2}),x(t_{1})\dot{x}(t_{1})+\dot{x}(t_{1})x(t_{1})]f(1)  \notag \\
&=&-m\dot{G}(t_{2}-t_{1})X(r;t_{1})+mG(t_{2}-t_{1})\dot{X}(r;t_{1}),
\label{6.35}
\end{eqnarray}
where we have used the relation (\ref{A.7}) to express the commutator in
terms of the Green function.. Also, we have 
\begin{equation}
\frac{\partial \dot{X}(r;t_{2})}{\partial r}=-m\ddot{G}
(t_{2}-t_{1})X(r;t_{1})+m\dot{G}(t_{2}-t_{1})\dot{X}(r;t_{1}).  \label{6.36}
\end{equation}
If we set $t_{2}=t_{1}$ and use the fact that $G(0)=0$ and $\dot{G}(0)=1/m$,
we find from (\ref{6.35}) that 
\begin{equation}
X(r;t_{1})=x(t_{1})e^{-r},  \label{6.37}
\end{equation}
while from (\ref{6.36}) we find that 
\begin{equation}
\dot{X}(r;t_{1})=\dot{x}(t_{1})e^{r}.  \label{6.38}
\end{equation}
Here we should emphasize that we use the single relaxation time model, for
which $\ddot{G}(0)=0$. For the Ohmic model there would be an extra term.
Putting these results in (\ref{6.35}) and (\ref{6.36}) and integrating, we
find 
\begin{eqnarray}
X(r;t_{2}) &=&x(t_{2})-(1-e^{-r})m\dot{G}x(t_{1})+(e^{r}-1)mG\dot{x}(t_{1}) 
\notag \\
\dot{X}(r;t_{2}) &=&\dot{x}(t_{2})-(1-e^{-r})m\ddot{G}x(t_{1})+(e^{r}-1)m
\dot{G}\dot{x}(t_{1}).  \label{6.39}
\end{eqnarray}
where $G=G(t_{2}-t_{1})$.

Since $X(r;t_{2})$ and $\dot{X}(r;t_{2})$ are linear in the operators $
x(t_{1})$ and $\dot{x}(t_{1})$, they have the Gaussian property and we can
use the identity (\ref{B.6}) to evaluate the expression (\ref{6.33}). With
this result the Wigner characteristic function (\ref{6.5}) can be written in
the form 
\begin{equation}
\tilde{W}(Q,P;t_{2})=\exp \{-\frac{A_{11}P^{2}+2A_{12}QP+A_{22}Q^{2}}{2\hbar
^{2}}\},  \label{6.40}
\end{equation}
where for this present example, 
\begin{subequations}
\begin{eqnarray}
A_{11} &=&\left\langle X^{2}(r;t_{2})\right\rangle  \notag \\
&=&[1-(1-e^{-r})m\dot{G}]^{2}\left\langle x^{2}\right\rangle
+(e^{r}-1)^{2}m^{2}G^{2}\left\langle \dot{x}^{2}\right\rangle  \notag \\
&&+(1-e^{-r})m\dot{G}s+(e^{r}-1)mG\dot{s},  \label{6.41a} \\
A_{12} &=&\frac{m}{2}\left\langle X(r;t_{2})\dot{X}(r;t_{2})+\dot{X}
(r;t_{2})X(r;t_{2})\right\rangle  \notag \\
&=&-[1-(1-e^{-r})m\dot{G}](1-e^{-r})m^{2}\ddot{G}\left\langle
x^{2}\right\rangle  \notag \\
&&+(e^{r}-1)^{2}m^{3}G\dot{G}\left\langle \dot{x}^{2}\right\rangle +\frac{1}{
2}(1-e^{-r})m^{2}\left( \ddot{G}s+\dot{G}\dot{s}\right)  \notag \\
&&+\frac{1}{2}(e^{r}-1)m^{2}\left( \dot{G}\dot{s}+G\ddot{s}\right) ,
\label{6.41b} \\
A_{22} &=&m^{2}\left\langle \dot{X}^{2}(r;t_{2})\right\rangle  \notag \\
&=&[1+(e^{r}-1)^{2}m^{2}\dot{G}^{2}]m^{2}\left\langle \dot{x}
^{2}\right\rangle +(1-e^{-r})^{2}m^{4}\ddot{G}^{2}\left\langle
x^{2}\right\rangle  \notag \\
&&.+(e^{r}-1)m^{3}\dot{G}\ddot{s}+(1-e^{-r})m^{3}\ddot{G}\dot{s}.
\label{6.41c}
\end{eqnarray}
Here we should again recall that $s=s(t_{2}-t_{1})$ is the mean square
displacement (\ref{A.8}) and $G=G(t_{2}-t_{1})$ is the Green function (\ref
{A.5}).

Forming the corresponding Wigner function, we find for the squeezed state, 
\end{subequations}
\begin{equation}
\mathcal{W}(q,p;t)=\frac{1}{2\pi \sqrt{A_{11}A_{22}-A_{12}^{2}}}\exp \{-
\frac{A_{11}p^{2}-2A_{12}pq+A_{22}q^{2}}{2(A_{11}A_{22}-A_{12}^{2})}.
\label{6.42}
\end{equation}
In Figure. \ref{qdist_fig4} we plot constant density contour for this
function in the plane of the dimensionless variables $u=q/\left\langle
x^{2}\right\rangle $ and $v=p/m\left\langle \dot{x}^{2}\right\rangle $. The
dashed circle corresponds to the equilibrium state, the state just before
the initial squeeze as well as the state at long times. The contour marked
(0) corresponds to the initial squeezed state. In the course of time this
contour rotates clockwise. The contour marked (1/4) is that corresponding to
a quarter period, while that marked (1/2) is that corresponding to a half
period when the squeezing is much reduced. The relatively strong coupling
chosen, $\gamma /\omega _{0}=10/13$, emphasizes the effect: dissipation
leads to a loss of squeezing.

\subsection{Example: \textquotedblleft Schr\"{o}dinger
cat\textquotedblright\ state}

Here we consider the Wigner function when the initial measurement
corresponds to the measurement function (\ref{5.17}). The calculation goes
exactly the same as that beginning with (\ref{5.5}) in the previous section,
so we shall simply quote the results. For general $f$ we find 
\begin{eqnarray}
\left\langle f^{\dag }(1)e^{-i(x(t_{2})P+m\dot{x}(t_{2})Q)/\hbar
}f(1)\right\rangle &=&\exp \{-\frac{\left\langle x^{2}\right\rangle
P^{2}+m^{2}\left\langle \dot{x}^{2}\right\rangle Q^{2}}{2\hbar ^{2}}+\frac{
(cP+m\dot{c}Q)^{2}}{2\left\langle x^{2}\right\rangle \hbar ^{2}}\}  \notag \\
&&\times \int_{-\infty }^{\infty }dx_{1}^{\prime }f^{\dag }(x_{1}^{\prime }+
\frac{GP+m\dot{G}Q}{2})f(x_{1}^{\prime }-\frac{GP+m\dot{G}Q}{2})  \notag \\
&&\times \frac{1}{\sqrt{2\pi \left\langle x^{2}\right\rangle }}\exp \{-\frac{
(x_{1}+x_{1}^{\prime })^{2}}{2\left\langle x^{2}\right\rangle }
-i(x_{1}+x_{1}^{\prime })\frac{cP+m\dot{c}Q}{\left\langle x^{2}\right\rangle
\hbar }\}.  \label{6.43}
\end{eqnarray}
Here $c=c(t_{2}-t_{1})$ is the correlation (\ref{A.9}) and $G=G(t_{2}-t_{1})$
is the Green function (\ref{A.5}). With the form (\ref{5.17}) for the
measurement function, this becomes (remember we must choose $x_{1}=0$ if we
wish the initial wave packet pair to be placed symmetrically about the
origin),
\begin{eqnarray}
&&\left\langle f^{\dag }(1)e^{-i(x(t_{2})P+m\dot{x}(t_{2})Q)/\hbar
}f(1)\right\rangle _{x_{1}=0}  \notag \\
&=&\frac{\exp \mathbf{\{}-\frac{A_{11}P^{2}+2A_{12}PQ+A_{22}Q^{2}}{2\hbar
^{2}}\}}{(1+e^{-d^{2}/8\sigma _{1}^{2}})\sqrt{2\pi (\left\langle
x^{2}\right\rangle +\sigma _{1}^{2})}}\left( e^{-d^{2}/8(\left\langle
x^{2}\right\rangle +\sigma _{1}^{2})}\cos \frac{(cP+m\dot{c}Q)d}{
2(\left\langle x^{2}\right\rangle +\sigma _{1}^{2})\hbar }\right.  \notag \\
&&\left. +e^{-d^{2}/8\sigma _{1}^{2}}\cosh \frac{(GP+m\dot{G}Q)d}{4\sigma
_{1}^{2}}\right) ,  \label{6.44}
\end{eqnarray}
where in this present example
\begin{eqnarray}
A_{11} &=&\left\langle x^{2}\right\rangle +\frac{\hbar ^{2}G^{2}}{4\sigma
_{1}^{2}}-\frac{c^{2}}{\left\langle x^{2}\right\rangle +\sigma _{1}^{2}}, 
\notag \\
A_{12} &=&m\left( \frac{\hbar ^{2}G\dot{G}}{4\sigma _{1}^{2}}-\frac{c\dot{c}
}{\left\langle x^{2}\right\rangle +\sigma _{1}^{2}}\right) ,  \notag \\
A_{22} &=&m^{2}\left( \left\langle \dot{x}^{2}\right\rangle +\frac{\hbar ^{2}
\dot{G}^{2}}{4\sigma _{1}^{2}}-\frac{\dot{c}^{2}}{\left\langle
x^{2}\right\rangle +\sigma _{1}^{2}}\right) .  \label{6.45}
\end{eqnarray}
The Wigner characteristic function is therefore
\begin{eqnarray}
\tilde{W}(Q,P;t) &=&\frac{\exp \{-\frac{A_{11}P^{2}+2A_{12}PQ+A_{22}Q^{2}}{
2\hbar ^{2}}\}}{1+\exp \{-\frac{\left\langle x^{2}\right\rangle d^{2}}{
8\sigma _{1}^{2}(\left\langle x^{2}\right\rangle +\sigma _{1}^{2})}\}}
\left\{ \cos \frac{(cP+m\dot{c}Q)d}{2(\left\langle x^{2}\right\rangle
+\sigma _{1}^{2})\hbar }\right.  \notag \\
&&\left. +\exp \{-\frac{\left\langle x^{2}\right\rangle d^{2}}{8\sigma
_{1}^{2}(\left\langle x^{2}\right\rangle +\sigma _{1}^{2})}\}\cosh \frac{
(GP+m\dot{G}Q)d}{4\sigma _{1}^{2}}\right\} .  \label{6.46}
\end{eqnarray}

With this form of the Wigner characteristic function there is no difficulty
evaluating the inverse transform (\ref{6.4}) to obtain the corresponding
Wigner function. However, the interest will be in the limits of a free
particle or, for the oscillator, a displaced ground state pair, in which
case it is simpler to first evaluate the limits of the above expression and
then evaluate the inverse transform.

\subsubsection{Free particle}

We obtain the case of a free particle coupled to the bath in the absence of
an oscillator potential by forming the limit $\left\langle
x^{2}\right\rangle \rightarrow \infty $. In forming this limit we should
recall that $c(t)=\left\langle x^{2}\right\rangle -s(t)/2$, where the mean
square displacement $s(t)$ is finite in the free particle limit. We find
\begin{eqnarray}
\tilde{W}(Q,P;t) &=&\frac{\exp \{-\frac{A_{11}P^{2}+2A_{12}PQ+A_{22}Q^{2}}{
2\hbar ^{2}}\}}{1+\exp \{-\frac{d^{2}}{8\sigma _{1}^{2}}\}}\left\{ \cos 
\frac{Pd}{2\hbar }\right.  \notag \\
&&\left. +\exp \{-\frac{d^{2}}{8\sigma _{1}^{2}}\}\cosh \frac{(GP+m\dot{G}Q)d
}{4\sigma _{1}^{2}}\right\} .  \label{6.47}
\end{eqnarray}
In this free particle case, the expressions (\ref{6.45}) become
\begin{eqnarray}
A_{11} &=&\sigma _{1}^{2}+s+\frac{\hbar ^{2}G^{2}}{4\sigma _{1}^{2}},  \notag
\\
A_{12} &=&m\left( \frac{\dot{s}}{2}+\frac{\hbar ^{2}G\dot{G}}{4\sigma
_{1}^{2}}\right) ,  \notag \\
A_{22} &=&m^{2}\left( \left\langle \dot{x}^{2}\right\rangle +\frac{\hbar ^{2}
\dot{G}^{2}}{4\sigma _{1}^{2}}\right) .  \label{6.48}
\end{eqnarray}
Forming the inverse Fourier transform (\ref{6.4}) we obtain
\begin{eqnarray}
\mathcal{W}(q,p;t) &=&\frac{1}{2\left( 1+e^{-d^{2}/8\sigma _{1}^{2}}\right) }
\left\{ \mathcal{W}_{0}(q-\frac{d}{2},p;t)+\mathcal{W}_{0}(q+\frac{d}{2}
,p;t)\right.  \notag \\
&&\left. +2\exp \{-A(t)\}\mathcal{W}_{0}(q,p;t)\mathbf{\cos }\Phi
(q,p;t)\right\} ,  \label{6.49}
\end{eqnarray}
where $\mathcal{W}_{0}$ is the Wigner function corresponding an initial
measurement forming a single wave packet at the origin, 
\begin{equation}
\mathcal{W}_{0}(q,p;t_{2})=\frac{\exp \{\mathbf{-}\frac{
A_{22}q^{2}-2A_{12}qp+A_{11}p^{2}}{2(A_{11}A_{22}-A_{12}^{2})}\}}{2\pi \sqrt{
A_{11}A_{22}-A_{12}^{2}}}.  \label{6.50}
\end{equation}
In this example the phase $\Phi $ is given by
\begin{equation}
\Phi (q,p;t)=\frac{(GA_{22}-m\dot{G}A_{12})q+(m\dot{G}A_{11}-GA_{12})p}{
A_{11}A_{22}-A_{12}^{2}}\frac{\hbar d}{4\sigma _{1}^{2}}  \label{6.51}
\end{equation}
and the quantity $A$ by
\begin{equation}
A(t)=\frac{(A_{11}-\frac{\hbar ^{2}G^{2}}{4\sigma _{1}^{2}})(A_{22}-\frac{
\hbar ^{2}m^{2}\dot{G}^{2}}{4\sigma _{1}^{2}})-(A_{12}-\frac{\hbar ^{2}mG
\dot{G}}{4\sigma _{1}^{2}})^{2}}{A_{11}A_{22}-A_{12}^{2}}\frac{d^{2}}{
8\sigma _{1}^{2}}.  \label{6.52}
\end{equation}

As in the case of the coherent state pair, The Wigner function for the free
particle \textquotedblleft Schr\"{o}dinger cat\textquotedblright\ state
shows three peaks, an outlying pair centered at $q=\pm d/2,p=0$ and an
interference peak centered at the origin. However, in this free particle
case 
\begin{equation}
A(0)=\frac{\sigma _{1}^{2}\left\langle \dot{x}^{2}\right\rangle }{\sigma
_{1}^{2}\left\langle \dot{x}^{2}\right\rangle +\frac{\hbar ^{2}}{4m^{2}}}
\frac{d^{2}}{8\sigma _{1}^{2}}=\frac{d^{2}}{8\sigma _{1}^{2}+2\bar{\lambda}
^{2}}>0  \label{6.53}
\end{equation}
where $\bar{\lambda}$ is the mean de Broglie wavelength in equilibrium,
\begin{equation}
\bar{\lambda}=\frac{\hbar }{m\sqrt{\left\langle \dot{x}^{2}\right\rangle }}.
\label{6.54}
\end{equation}
Therefore, the amplitude of the interference peak will initially be
vanishingly small whenever the separation $d$ of the wave packets is large
compared with both the slit width $\sigma _{1}$ and the mean de Broglie
wavelength $\bar{\lambda}=\hbar /m\sqrt{\left\langle \dot{x}
^{2}\right\rangle }$.

\subsubsection{Displaced ground state pair}

Again we form the limit $\sigma _{1}^{2}\rightarrow \infty $ and $
d\rightarrow \infty $ such that $d_{0}=\left\langle x^{2}\right\rangle
d/\sigma _{1}^{2}$ is fixed. The coefficients (\ref{6.45}) become
\begin{equation}
A_{11}=\left\langle x^{2}\right\rangle ,\quad A_{12}=0,\quad
A_{22}=m^{2}\left\langle \dot{x}^{2}\right\rangle ..  \label{6.55}
\end{equation}
The Wigner characteristic function (\ref{6.46}) then becomes
\begin{eqnarray}
\tilde{W}(Q,P;t) &=&\frac{\exp \{-\frac{\left\langle x^{2}\right\rangle
P^{2}+m^{2}\left\langle \dot{x}^{2}\right\rangle Q^{2}}{2\hbar ^{2}}\}}{
1+\exp \{-d_{0}^{2}/8\left\langle x^{2}\right\rangle \}}\left\{ \cos \frac{
(cP+m\dot{c}Q)d_{0}}{2\left\langle x^{2}\right\rangle \hbar }\right.  \notag
\\
&&\left. +\exp \{-d_{0}^{2}/8\left\langle x^{2}\right\rangle \}\cosh \frac{
(GP+m\dot{G}Q)d_{0}}{4\left\langle x^{2}\right\rangle }\right\} .
\label{6.56}
\end{eqnarray}
With this, the Wigner function is
\begin{eqnarray}
\mathcal{W}(q,p;t) &=&\frac{1}{2\left( 1+\exp \{-\frac{d_{0}^{2}}{
8\left\langle x^{2}\right\rangle }\}\right) }\left\{ \mathcal{W}_{\text{eq}
}(q-\frac{cd_{0}}{2\left\langle x^{2}\right\rangle },p-\frac{m\dot{c}d_{0}}{
2\left\langle x^{2}\right\rangle })\right.  \notag \\
&&\left. +\mathcal{W}_{\text{eq}}(q+\frac{cd_{0}}{2\left\langle
x^{2}\right\rangle },p+\frac{m\dot{c}d_{0}}{2\left\langle x^{2}\right\rangle 
})+2e^{-A}\mathcal{W}_{\text{eq}}(q,p)\cos \Phi \right\}  \label{6.57}
\end{eqnarray}
where $\mathcal{W}_{\text{eq}}(q,p)$ is the equilibrium Wigner function (\ref
{6.8}) and
\begin{eqnarray}
\Phi (q,p;t) &=&\left( \frac{\hbar G}{\left\langle x^{2}\right\rangle }q+
\frac{\hbar \dot{G}}{m\left\langle \dot{x}^{2}\right\rangle }p\right) \frac{
d_{0}}{4\left\langle x^{2}\right\rangle },  \notag \\
A(t) &=&\frac{d_{0}^{2}}{8\left\langle x^{2}\right\rangle }\left( 1-\frac{
\hbar ^{2}G^{2}}{4\left\langle x^{2}\right\rangle ^{2}}-\frac{\hbar ^{2}\dot{
G}^{2}}{4\left\langle x^{2}\right\rangle \left\langle \dot{x}
^{2}\right\rangle }\right) .  \label{6.58}
\end{eqnarray}
Again, as in the free particle case the initial value of $A$ is not zero,
\begin{equation}
A(0)=\frac{d_{0}^{2}}{8\left\langle x^{2}\right\rangle }\left( 1-\frac{\hbar
^{2}}{4m^{2}\left\langle \dot{x}^{2}\right\rangle \left\langle
x^{2}\right\rangle }\right) >0,  \label{6.59}
\end{equation}
and the interference term is vanishingly small.

This situation, in which the initial state is \ a \textquotedblleft Schr\"{o}
dinger Cat\textquotedblright\ state formed by passing the particle through a
pair of Gaussian slits, is to be contrasted with that described in Sec. \ref
{sec:six-C}, in which the initial state is prepared by displacing the
equilibrium state to form a coherent state pair. The Wigner functions, given
by Eqs. (\ref{6.58}) and (\ref{6.22}), respectively, are identical in form,
but in the \textquotedblleft Schr\"{o}dinger Cat\textquotedblright\ case the
interference peak is initially vanishingly small and remains so for all
time, while in that of the coherent state pair the interference peak is
initially twice as high as the outlying peaks, becoming vanishingly small
only after a short relaxation time. The reverse is true for the probability
distributions, given in Eqs. (\ref{5.24}) and (\ref{6.26}), respectively.
That is, the interference term in the probability distribution is
vanishingly small at all times for the coherent state pair, while for the
\textquotedblleft Schr\"{o}dinger Cat\textquotedblright\ state the
attenuation coefficient multiplying the interference term vanishes only
after a short decoherence. time. We must conclude that the notion of
decoherence. time is arbitrary, depending on the situation and how one
chooses to view it.

\section{Concluding remarks}

\label{sec:seven}

The quantum probability distributions are \emph{measured} distributions.
That is, they depend explicitly on the parameters of the measurements. This
is seen clearly in the general formula (\ref{3.29}) for the n-point
distribution, where $f(j)$ is the measurement function for the $j$'th
measurement. This is also seen in the in the case of quantum Brownian motion
in the expression (\ref{4.2}) for the characteristic function. There it is
seen that the characteristic function can be factored, with a factor $
K(1,\cdots ,n)$ multiplying a quantum expectation independent of
measurement. The factor $K(1,\cdots ,n)$ contains the measurement parameters
and depends upon the dynamics through the non-equal-time commutator. In the
classical limit, where this commutator vanishes, this factor becomes a
numerical factor independent of the dynamics that can in practice be taken
to be unity. The result is the familiar expression for the classical
characteristic function.\cite{doob}

An important part of this work has been to demonstrate that the general
expression (\ref{4.2}) for the characteristic function can be very useful
for practical calculations. The use of this formula together with its
specializations (\ref{5.9}) to the probability distribution and (\ref{6.5})
to the Wigner function has been illustrated with a number of examples, each
of which is an important application. Among the results we point out the
expression (\ref{5.13}) for wave packet spreading in the presence of
dissipation, a generalization of the well known expression found in
elementary textbooks. \ Another result is illustrated in Fig. \ref
{qdist_fig4} where the disappearance of squeezing in the presence of
dissipation is illustrated. Finally, a comparison of a \textquotedblleft Schr
\"{o}dinger Cat\textquotedblright\ state formed by passing the particle
through a pair of Gaussian slits with the nearly identical state formed with
a coherent state pair shows that a quantitative measure of decoherence.
depends on how the state is formed.

In Sec. \ref{sec:six} we discuss the Wigner function. Some authors prefer
instead the density matrix element in the coordinate representation, which
is given by a kind of half-Fourier transform:
\begin{equation}
\left\langle x\left\vert \rho (t)\right\vert x^{\prime }\right\rangle
=\int_{-\infty }^{\infty }dp\mathcal{W}(\frac{x+x^{\prime }}{2}
,p;t)e^{i(x-x^{\prime })p/\hbar }.  \label{7.1}
\end{equation}
Clearly, the Wigner function and the density matrix element contain the same
information. Indeed, it is not difficult to see that the central
interference peak in, say, the Wigner function (\ref{6.49}) corresponding to
a coherent state pair becomes a pair of off-diagonal peaks in the density
matrix element. We prefer the language of the Wigner function since it is
always a real function that in the classical limit becomes a real
probability distribution.

The examples are all in one dimension. The reader should be aware that this
is not a necessary restriction, but has been made to keep the discussion
within bounds. The generalization to higher dimensions is straightforward.
All that one must keep in mind is that the fluctuating force operator is
independent in the different directions and that the effects of dissipation
(the fluctuating force \ and the memory force) are independent of the
applied force.\cite{ford88b} Finally, we have restricted the discussion to
the single relaxation time model of dissipation and its limiting Ohmic case.
Again this has been done to keep the discussion within bounds. There is no
problem with the discussion for more general models such as the coupling to
the blackbody radiation field.\cite{ford88b}

\appendix

\section{Quantum Brownian motion}

\label{appendix:A}

\subsection{Quantum Langevin equation}

Quantum Brownian motion for an oscillator coupled to a heat bath at
temperature $T$ is described by the quantum Langevin equation, 
\begin{equation}
m\ddot{x}+\int_{-\infty }^{t}dt^{\prime }\mu (t-t^{\prime })\dot{x}
(t^{\prime })+Kx=F(t).  \label{A.1}
\end{equation}
This is a Heisenberg equation for the position operator $x(t)$. On the right
hand side $F(t)$ is a Gaussian random operator force, with mean zero, $
\left\langle F(t)\right\rangle =0$, and with autocorrelation and commutator
given by 
\begin{eqnarray}
\frac{1}{2}\left\langle F(t)F(t^{\prime })+F(t^{\prime })F(t)\right\rangle
&=&\frac{\hbar }{\pi }\int_{0}^{\infty }d\omega \mathrm{Re}\{\tilde{\mu}
(\omega +i0^{+})\}\omega \coth \frac{\hbar \omega }{2kT}\cos \omega
(t-t^{\prime }),  \notag \\
\lbrack F(t),F(t^{\prime })] &=&\frac{\hbar }{i\pi }\int_{0}^{\infty
}d\omega \mathrm{Re}\{\tilde{\mu}(\omega +i0^{+})\}\omega \sin \omega
(t-t^{\prime }).  \label{A.2}
\end{eqnarray}
In these expressions $\tilde{\mu}$ is the Fourier transform of the memory
function, 
\begin{equation}
\tilde{\mu}(z)=\int_{0}^{\infty }dt\mu (t)e^{izt},\qquad \mathrm{Im}\{z\}>0.
\label{A.3}
\end{equation}
It is a consequence of the second law of thermodynamics that $\tilde{\mu}(z)$
must be what is called a positive real function: analytic with real part
positive everywhere in the upper half plane.

In our present discussion we take the view that the above is a macroscopic
description, which is complete as it stands. For a thorough discussion,
including the derivation from a number of microscopic models, we refer to a
paper of Ford, Lewis and O'Connell \cite{ford88b}.

The solution of the quantum Langevin equation (\ref{A.1}) can be written 
\begin{equation}
x(t)=\int_{-\infty }^{t}dt^{\prime }G(t-t^{\prime })F(t^{\prime }),
\label{A.4}
\end{equation}
where the Green function $G(t)$ is given by 
\begin{equation}
G(t)=\frac{1}{2\pi }\int_{-\infty }^{\infty }d\omega \alpha (\omega
+i0^{+})e^{-i\omega t},  \label{A.5}
\end{equation}
in which $\alpha (z)$, the response function, is given by 
\begin{equation}
\alpha (z)=\frac{1}{-mz^{2}-iz\tilde{\mu}(z)+K}.  \label{A.6}
\end{equation}

With this solution, we can obtain the following expressions for the
commutator,
\begin{eqnarray}
\lbrack x(t),x(t^{\prime })] &=&\frac{2\hbar }{i\pi }\int_{0}^{\infty
}d\omega \mathrm{Im}\{\alpha (\omega +i0^{+})\}\sin \omega (t-t^{\prime }) 
\notag \\
&=&i\hbar \{G(t^{\prime }-t)-G(t-t^{\prime })\}.  \label{A.7}
\end{eqnarray}
The Green function vanishes for negative times, while $G(0)=0$ and $\dot{G}
(0)=1/m$. We see therefore, that the canonical commutator, $[x,p]=i\hbar $,
holds with $p=m\dot{x}$, the mechanical momentum. (The canonical momentum
may or may not be the same as the mechanical momentum, depending on the form
of the microscopic Hamiltonian.)

Also of interest is the mean square displacement,
\begin{eqnarray}
s(t-t^{\prime }) &\equiv &\left\langle (x(t)-x(t^{\prime }))^{2}\right\rangle
\notag \\
&=&2\left\langle x^{2}\right\rangle -2c(t-t^{\prime }).  \label{A.8}
\end{eqnarray}
Here $c(t)$ is the correlation
\begin{equation}
c(t)=\frac{1}{2}\left\langle x(t)x(0)+x(0)x(t)\right\rangle .  \label{A.9}
\end{equation}
Using the solution (\ref{A.4}) or, more directly without recourse to the
Langevin equation, the fluctuation-dissipation theorem of Callen and Welton 
\cite{callen51}, we obtain the following expression for the correlation
\begin{equation}
c(t)=\frac{\hbar }{\pi }\int_{0}^{\infty }d\omega \mathrm{Im}\{\alpha
(\omega +i0^{+})\}\coth \frac{\hbar \omega }{2kT}\cos \omega t.  \label{A.10}
\end{equation}
With this, we also have
\begin{equation}
s(t)=\frac{2\hbar }{\pi }\int_{0}^{\infty }d\omega \mathrm{Im}\{\alpha
(\omega +i0^{+})\}\coth \frac{\hbar \omega }{2kT}(1-\cos \omega t).
\label{A.11}
\end{equation}
Here we should note first of all that for very long times the correlation
vanishes and
\begin{equation}
s(t)\sim 2\left\langle x^{2}\right\rangle .  \label{A.12}
\end{equation}
The exception is the free particle case, where $K=0$ and consequently $
\left\langle x^{2}\right\rangle =\infty $. In that case $s(t)$ grows for
long times without limit, with a time dependence that depends on the model
as well as the temperature \cite{grabert87,ford06c}. On the other hand, for
very short times we can expand the cosine to obtain the general result
\begin{equation}
s(t)\cong \left\langle \dot{x}^{2}\right\rangle t^{2}.  \label{A.13}
\end{equation}

\subsection{Explicit expressions}

Here we obtain explicit, closed form expressions for the Green function $
G(t) $ and the mean square displacement $s(t)$. For this purpose, the model
of choice for most applications, due to its simplicity, is the Ohmic model,
for which $\tilde{\mu}(z)=\zeta $, the friction constant. While it is
adequate for most purposes, this Ohmic model is singular, particularly at
short times or high frequencies, so we here consider a more general model in
which the higher frequencies are suppressed. The simplest of these is the
single relaxation time model, for which
\begin{equation}
\tilde{\mu}(z)=\frac{\zeta }{1-iz\tau }.  \label{A.14}
\end{equation}
Here $\tau $ is the relaxation time (at times called the bath correlation
time) which we assume is small in the sense that $\zeta \tau /m\ll 1$.
Putting this form of $\tilde{\mu}$ in the response function (\ref{A.6}) and
replacing the parameters $\tau $, $\zeta $ and $K$ with parameters $\Omega $
, $\gamma $ and $\omega _{0}$ through the relations
\begin{equation}
\tau =\frac{1}{\Omega +\gamma },\quad \zeta =m\gamma \frac{\Omega (\Omega
+\gamma )+\omega _{0}^{2}}{(\Omega +\gamma )^{2}},\quad K=m\omega _{0}^{2}
\frac{\Omega }{\Omega +\gamma },  \label{A.15}
\end{equation}
we obtain the convenient form
\begin{equation}
\alpha (z)=\frac{z+i(\Omega +\gamma )}{m(z+i\Omega )(-z^{2}-i\gamma z+\omega
_{0}^{2})}.  \label{A.16}
\end{equation}
Note that $-i\Omega $ is the pole of the response function far down on the
negative imaginary axis. For small relaxation time $\Omega \sim 1/\tau $ we
have the expansion
\begin{equation}
\Omega =\frac{1}{\tau }-\frac{\zeta }{m}-\frac{\zeta ^{2}\tau }{m^{2}}+(
\frac{K\zeta }{m^{2}}-2\frac{\zeta ^{3}}{m^{3}})\tau ^{2}+\cdots .
\label{A.17}
\end{equation}
The Ohmic model, for which $\tau \rightarrow 0$, therefore corresponds to $
\Omega \rightarrow \infty $, while $\gamma \rightarrow \zeta /m$ and $\omega
_{0}^{2}\rightarrow K/m$.

\subsubsection{Green function}

With the form (\ref{A.16}) of the response function, we evaluate the
integral in the expression (\ref{A.5}) for the Green function by deforming
the path of integration into the lower half plane, picking up the residues
at the poles of the response function. The result is
\begin{eqnarray}
G(t) &=&\frac{\gamma }{m(\Omega ^{2}-\gamma \Omega +\omega _{0}^{2})}
(e^{-\Omega t}-e^{-\gamma t/2}\cos \omega _{1}t)  \notag \\
&&+\frac{\Omega ^{2}+\omega _{0}^{2}-\gamma ^{2}/2}{\Omega ^{2}-\gamma
\Omega +\omega _{0}^{2}}e^{-\gamma t/2}\frac{\sin \omega _{1}t}{m\omega _{1}}
,  \label{A.18}
\end{eqnarray}
where
\begin{equation}
\omega _{1}=\sqrt{\omega _{0}^{2}-\gamma ^{2}/4}.  \label{A.19}
\end{equation}
In the limit, $\Omega \rightarrow \infty $, this becomes the familiar Ohmic
Green function
\begin{equation}
G_{\mathrm{Ohmic}}(t)=e^{-\gamma t/2}\frac{\sin \omega _{1}t}{m\omega _{1}}.
\label{A.20}
\end{equation}

For reasonable choices of the parameters the Ohmic Green function is very
little different from that for the single relaxation time model. The
difference between the two models is more apparent in the second derivative
of the Green function, which we show in Figure \ref{qdist_fig5} for the two
models. The difference is small except for short times, where $\ddot{G}(0)$
vanishes, while $\ddot{G}_{\mathrm{Ohmic}}(0)=-\zeta /m^{2}$ is finite.

\subsubsection{Mean square displacement at high temperature}

First we consider the mean square displacement in the high temperature
limit, in which in the expression (\ref{A.10}) we replace the hyperbolic
cotangent by the reciprocal of its argument. If we compare the resulting
expression with the expression (\ref{A.5}) for the Green function, we see
that in this high temperature limit,
\begin{equation}
s(t)=2kT\int_{0}^{t}dt^{\prime }G(t^{\prime }).  \label{A.21}
\end{equation}
The long and short time behavior of the mean square displacement are of the
general form given in Eqs. (\ref{A.12}) and (\ref{A.13}). Note that in this
high temperature limit $\left\langle x^{2}\right\rangle $ and $\left\langle 
\dot{x}^{2}\right\rangle $ are given by the classical equipartition formulas,
\begin{equation}
\left\langle x^{2}\right\rangle =\frac{kT}{K},\qquad \left\langle \dot{x}
^{2}\right\rangle =\frac{kT}{m}.  \label{A.22}
\end{equation}

\subsubsection{Mean square displacement at zero temperature}

We consider now the effect of zero-point oscillations on the mean square
displacement. At temperature zero, we replace the hyperbolic cotangent in
the expression (\ref{A.10}) by unity. Then if we write,
\begin{eqnarray}
\mathrm{Im}\{\alpha (\omega +i0^{+})\} &=&-\frac{\gamma }{m(\Omega
^{2}-\gamma \Omega +\omega _{0}^{2})}\frac{\Omega ^{2}}{\omega (\omega
^{2}+\Omega ^{2})}  \notag \\
&&+\mathrm{Im}\left\{ \frac{\Omega ^{2}-(\frac{\gamma }{2}-i\omega _{1})^{2}
}{m\omega _{1}(\Omega ^{2}-\gamma \Omega +\omega _{0}^{2})}\frac{(\frac{
\gamma }{2}+i\omega _{1})^{2}}{\omega \lbrack \omega ^{2}+(\frac{\gamma }{2}
+i\omega _{1})^{2}]}\right\} ,  \label{A.23}
\end{eqnarray}
we can write the expression (\ref{A.11}) for the mean square displacement in
the form
\begin{equation}
s(t)=\frac{2\hbar }{m\pi }\left( -\frac{\gamma }{\Omega ^{2}-\gamma \Omega
+\omega _{0}^{2}}V(\Omega t)+\mathrm{Im}\left\{ \frac{\Omega ^{2}-(\frac{
\gamma }{2}-i\omega _{1})^{2}}{\omega _{1}(\Omega ^{2}-\gamma \Omega +\omega
_{0}^{2})}V(\frac{\gamma t}{2}+i\omega _{1}t)\right\} \right) .  \label{A.24}
\end{equation}
Here we have introduced the function ,
\begin{eqnarray}
V(z) &\equiv &\int_{0}^{\infty }dy\frac{1-\cos zy}{y(1+y^{2})}  \notag \\
&=&\log z+\gamma _{\mathrm{E}}-\frac{1}{2}[e^{-z}\mathrm{Ei}(z)+e^{z}\mathrm{
Ei}(-z)],  \label{A.25}
\end{eqnarray}
where $\gamma _{\mathrm{E}}=0.577215665$ is Euler's constant and $\mathrm{Ei}
$ is the exponential integral \cite{bateman_htf2} . Using the expansion of
the exponential integral for small argument, we obtain the expansion
\begin{equation}
V(z)=-(\log z+\gamma _{\mathrm{E}})(\cosh z-1)-\frac{1}{2}
[e^{-z}\sum_{n=1}^{\infty }\frac{z^{n}}{n!n}+e^{z}\sum_{n=1}^{\infty }\frac{
(-z)^{n}}{n!n}].  \label{A.26}
\end{equation}
We see from this that for small $z$, 
\begin{equation}
V(z)\cong -\frac{1}{2}z^{2}(\log z+\gamma _{\mathrm{E}}-\frac{3}{2})-\frac{1
}{24}z^{4}(\log z+\gamma _{\mathrm{E}}-\frac{25}{12})+\cdots .  \label{A.27}
\end{equation}
Note that $V(0)=0$, in agreement with the definition (\ref{A.25}). For large 
$z$, using the asymptotic formulas for the exponential integral, we obtain
the asymptotic expansion, 
\begin{equation}
V(z)\sim \log z+\gamma _{\mathrm{E}}-\frac{1}{z^{2}}-\frac{3!}{z^{4}}-\frac{
5!}{z^{6}}-\cdots .  \label{A.28}
\end{equation}

With these results, we see that for very short times ($t\ll \tau $) the mean
square displacement again takes the form (\ref{A.13}) but now with the mean
square velocity, given by
\begin{eqnarray}
\left\langle \dot{x}^{2}\right\rangle &=&\hbar \frac{\lbrack \Omega
^{2}(\omega _{0}^{2}-\frac{\gamma ^{2}}{2})+\omega _{0}^{4}]\arccos \frac{
\gamma }{2\omega _{0}}+\gamma \omega _{1}\Omega ^{2}\log \frac{\Omega }{
\omega _{0}}}{\pi m\omega _{1}(\Omega ^{2}-\gamma \Omega +\omega _{0}^{2})} 
\notag \\
&\cong &\frac{\hbar }{\pi m}\{-\gamma \log \omega _{0}\tau +\frac{\omega
_{0}^{2}-\frac{\gamma ^{2}}{2}}{\omega _{1}}\arccos \frac{\gamma }{2\omega
_{0}}\},  \label{A.29}
\end{eqnarray}
where second form is that for small relaxation time, $\Omega \rightarrow
1/\tau $. In the Ohmic limit this is logarithmically divergent, so we have
here a case where the single relaxation time makes a difference. On the
other hand, 
\begin{eqnarray}
\left\langle x^{2}\right\rangle &=&\hbar \frac{(\Omega ^{2}+\omega _{0}^{2}-
\frac{\gamma ^{2}}{2})\arccos \frac{\gamma }{2\omega _{0}}-\gamma \omega
_{1}\log \frac{\Omega }{\omega _{0}}}{\pi m\omega _{1}(\Omega ^{2}-\gamma
\Omega +\omega _{0}^{2})}  \notag \\
&\cong &\frac{\hbar }{\pi m\omega _{1}}\arccos \frac{\gamma }{2\omega _{0}}.
\label{A.30}
\end{eqnarray}
Thus the Ohmic limit of $\left\langle x^{2}\right\rangle $ is finite.

\subsubsection{Free particle}

The free particle corresponds to the absence of the oscillator potential,
that is to the limit $K\rightarrow 0$. For the Green function we can obtain
this limit by setting $\omega _{0}=0$ and $\omega _{1}=i\gamma /2$ in the
expression (\ref{A.18}). This gives
\begin{equation}
G(t)=\frac{\Omega ^{2}(1-e^{-\gamma t})-\gamma ^{2}(1-e^{-\Omega t})}{\zeta
(\Omega ^{2}-\gamma ^{2})}.  \label{A.31}
\end{equation}
In this free particle case the parameters $\Omega $ and $\gamma $ are given
by the relations (\ref{A.15}) with $K=0$. These can then be inverted to give
\begin{equation}
\Omega =\frac{1+\sqrt{1-4\zeta \tau /m}}{2\tau },\quad \gamma =\frac{1-\sqrt{
1-4\zeta \tau /m}}{2\tau }.  \label{A.32}
\end{equation}

With this expression for the Green function, the high temperature form (\ref
{A.21}) of the mean square displacement becomes
\begin{eqnarray}
s(t) &=&\frac{2kT}{\zeta }\{t-\frac{\Omega ^{3}(1-e^{-\gamma t})-\gamma
^{3}(1-e^{-\Omega t})}{\gamma \Omega (\Omega ^{2}-\gamma ^{2})}\}  \notag \\
&\cong &\frac{2kT}{\zeta }(t-\frac{1-e^{-\gamma t}}{\gamma }).  \label{A.33}
\end{eqnarray}
Note that at long time this increases linearly with time, consistent with (
\ref{A.12}) in the sense that $\left\langle x^{2}\right\rangle =\infty $ for
the free particle. On the other hand for short times, we get exactly the
short time result (\ref{A.13}) with $\left\langle \dot{x}^{2}\right\rangle $
given by the equipartition form (\ref{A.22}). In other words, the short time
behavior of the oscillator is that of the free particle.

At zero temperature the result (\ref{A.24}) becomes for the free particle,
\begin{equation}
s(t)=\frac{2\hbar }{\pi \zeta }\frac{\Omega ^{2}V(\gamma t)-\gamma
^{2}V(\Omega t)}{\Omega ^{2}-\gamma ^{2}}.  \label{A.34}
\end{equation}
At very short times ($t\ll \tau $) this takes the form (\ref{A.13}) with now
\begin{equation}
\left\langle \dot{x}^{2}\right\rangle =\frac{\hbar \gamma \Omega }{\pi
m(\Omega -\gamma )}\log \frac{\Omega }{\gamma }\cong -\frac{\hbar \gamma }{
\pi m}\log \gamma \tau .  \label{A.35}
\end{equation}
At very long times ($t\gg \gamma ^{-1}$), we find
\begin{eqnarray}
s(t) &\sim &\frac{2\hbar }{\pi \zeta }\{\frac{\Omega +\gamma }{\Omega }(\log
\gamma t+\gamma _{\mathrm{E}})-\frac{\gamma ^{2}}{\Omega (\Omega -\gamma )}
\log \frac{\Omega }{\gamma }\}  \notag \\
&\cong &\frac{2\hbar }{\pi \zeta }\log \zeta t/m.  \label{A.36}
\end{eqnarray}

\subsection{Weak coupling}

The coupling to the heat bath is measured by the function $\tilde{\mu}(z)$.
If this is small, the response function will be sharply peaked about $\omega
_{0}=\sqrt{K/m}$, the natural frequency of the oscillator. In the integral
expression (\ref{A.5}) for the Green function, one is therefore led to make
the replacement $\tilde{\mu}(\omega )\rightarrow \tilde{\mu}(\omega _{0})$.
The result is an expression of the Ohmic form (\ref{A.20}) with
\begin{equation}
\gamma =\frac{1}{m}\mathrm{Re}\{\tilde{\mu}(\omega _{0})\}.  \label{A.37}
\end{equation}
(The imaginary part gives a negligible contribution to $\omega _{0}$.) Next,
in the integral expression (\ref{A.11}) for the mean square displacement, we
make the same approximations with, in addition the replacement $\hbar \omega
\coth \frac{\hbar \omega }{2kT}\rightarrow \hbar \omega _{0}\coth \frac{
\hbar \omega _{0}}{2kT}$, to obtain
\begin{eqnarray}
s(t) &\cong &2m\left\langle \dot{x}^{2}\right\rangle \int_{0}^{t}dt^{\prime
}G(t^{\prime })  \notag \\
&=&2\left\langle x^{2}\right\rangle \{1-e^{-\gamma t/2}(\cos \omega _{1}t+
\frac{\gamma }{2\omega _{1}}\sin \omega _{1}t)\},  \label{A.38}
\end{eqnarray}
where
\begin{equation}
\left\langle \dot{x}^{2}\right\rangle \cong \omega _{0}^{2}\left\langle
x^{2}\right\rangle \cong \frac{\hbar \omega _{0}}{2m}\coth \frac{\hbar
\omega _{0}}{2kT}.  \label{A.39}
\end{equation}

This constitutes the weak coupling approximation \cite{ford99}. We should
emphasize that this weak coupling approximation is not valid for the free
particle, as should be clear from the above argument. Note, incidentally
that for the Ohmic model this approximation is exact in the high temperature
limit. The usual statement is that it is valid in the limit $\gamma \ll
\omega _{0}$, but consideration of the exact results given above tells us
that even in this limit the approximation is not correct for very short
times ($\omega _{0}t\ll 1$) nor for very long times ($\gamma t\gg 1$).
However, for all other times, the weak coupling approximation is very good
for surprisingly large values of the coupling. As an illustration in Figure 
\ref{qdist_fig6} we compare the mean square displacement at zero temperature
as calculated first with the exact formula (\ref{A.24}) and then with the
weak coupling approximation. The parameters $\Omega /\omega _{0}=5$ and $
\gamma /\omega _{0}=10/13$ were chosen to exaggerate the difference. What we
see is that the weak coupling approximation is surprisingly good, even for
rather strong coupling.

If one makes the further approximation of neglecting quantities of relative
order $\gamma /\omega _{0}$, one gets what at times is called the
Weisskopf-Wigner approximation. In Eq. (\ref{A.38}\} it would correspond to
replacing $\omega _{1}\rightarrow \omega _{0}$ and dropping the second term
after the exponential. This then would correspond exactly to what one
obtains by solving the well known weak coupling master equation \cite
{louisell64}. Because of this, in the literature the Weisskopf-Wigner
approximation is often called the weak coupling approximation. The
difference between the weak coupling approximations we have defined it and
the Weisskopf-Wigner approximation is illustrated dramatically in Fig. \ref
{qdist_fig3}.

\section{Mathematical formulas}

\label{appendix:B}

Here we collect some formulas used in the evaluation of the various
examples. These formulas are all simple and more or less well known. The
first is the standard Gaussian integral,
\begin{equation}
\int_{-\infty }^{\infty }du\exp \{-\frac{1}{2}au^{2}+bu\}=\sqrt{\frac{2\pi }{
a}}\exp \{\frac{b^{2}}{2a}\}.  \label{B.1}
\end{equation}
The generalization to $d$ dimensions takes the form
\begin{equation}
\int d\mathbf{u}\exp (-\frac{1}{2}\mathbf{u}\cdot \mathbf{A}\cdot \mathbf{u}+
\mathbf{B}\cdot \mathbf{u}\}=\frac{(2\pi )^{d/2}}{\sqrt{\det \mathbf{A}}}
\exp \{\frac{1}{2}\mathbf{B}\cdot \mathbf{A}^{-1}\cdot \mathbf{B}\}.
\label{B.2}
\end{equation}
Here we have used dyadic notation, with $\mathbf{A}$ a positive definite
symmetric matrix and $\mathbf{B}$ a vector (not necessarily real) in $d$
dimensions. This generalization follows from the standard integral, using
the fact that a symmetric matrix can be diagonalized by an orthogonal
transformation.

The second formula is the Baker-Campbell-Hausdorf formula. If $A$ and $B$
are a pair of operators (not necessarily Hermitian) whose commutator is a
c-number, then
\begin{equation}
e^{A}e^{B}=e^{A+B}e^{[A,B]/2}=e^{B}e^{A}e^{[A,B]}.  \label{B.3}
\end{equation}
This formula is easily checked by expanding the exponentials in powers of
their argument. A generalization of this theorem is the convenient formula
\begin{equation}
e^{A}g(B)=g(B+[A,B])e^{A},  \label{B.4}
\end{equation}
which holds for a general function $g(B)$. Again, this can be verified by
expanding $g$ in powers of its argument.

Finally, we have a couple of formulas based on the notion of a Gaussian
variable. In general a set of operators (each with mean zero) is Gaussian if
the expectation of a product of an odd number of the operators is zero while
the product of an even number is equal to the sum of products of pair
expectations, the sum being over all $(2n-1)!!$ pairings with the order
within the pairs preserved. A Gaussian variable, e.g. $x(t)$, is such a set
with the members labeled with the time. Thus, for example, with an obvious
shorthand,
\begin{equation}
\left\langle 1234\right\rangle =\left\langle 12\right\rangle \left\langle
34\right\rangle +\left\langle 13\right\rangle \left\langle 24\right\rangle
+\left\langle 14\right\rangle \left\langle 23\right\rangle .  \label{B.5}
\end{equation}
Note that within each pair the order is the same as the original order. A
straightforward consequence of this Gaussian property is that for a Gaussian
operator $\mathcal{O}$, we have the formula
\begin{equation}
\left\langle e^{i\mathcal{O}}\right\rangle =e^{-\frac{1}{2}\left\langle 
\mathcal{O}^{2}\right\rangle }.  \label{B.6}
\end{equation}
This is easily verifies by expanding the exponentials. Another convenient
result is
\begin{equation}
\left\langle \frac{1}{\sqrt{2\pi \sigma ^{2}}}\exp \{-\frac{(\mathcal{O}
-a)^{2}}{2\sigma ^{2}}\}\right\rangle =\frac{1}{\sqrt{2\pi (\sigma
^{2}+\left\langle \mathcal{O}^{2}\right\rangle )}}\exp \{-\frac{a^{2}}{
2(\sigma ^{2}+\left\langle \mathcal{O}^{2}\right\rangle )}\}.  \label{B.7}
\end{equation}
To obtain this result, form the Fourier transform with respect to the
parameter $a$ using the standard Gaussian integral (\ref{B.1}) and then form
the expectation using (\ref{B.6}).

\begin{center}
{\Large Figure Captions}
\end{center}

Figure 1. Probability distribution for a free particle in the Schr\"{o}
dinger cat state. $P(x,0)$ is the initial distribution. $P(x,t)$ is the
distribution at time $t$, while $P_{0}(x,t)$ is the distribution obtained at
time $t$ by artificially setting the attenuation coefficient equal to zero.

Figure 2. The motion of the wave packet center for the displaced ground
state and for the coherent state, both for initial velocity zero. The
displaced ground state motion is computed at zero temperature. The
parameters chosen are $\gamma /\omega _{0}=10/13$ and $\Omega /\omega _{0}=5$
.

Figure 3. The function $A(t)$ for a coherent state pair. The parameters
chosen are $\gamma /\omega _{0}=10/13$, $\Omega /\omega _{0}=5$.

Figure 4. Constant density contours of the Wigner function for a squeezed
state, shown in the plane of the dimensionless variables $u=q/\sqrt{
\left\langle x^{2}\right\rangle }$ and $v=p/m\sqrt{\left\langle \dot{x}
^{2}\right\rangle }$. The dashed circle corresponds to the equilibrium
state, the state just before the initial squeeze as well as the state at
long times. The contour marked (0) corresponds to the initial squeezed
state. The contour marked (1/4) is that corresponding to a quarter period,
while that marked (1/2) is that corresponding to a half period. The
parameters chosen are $\gamma /\omega _{0}=10/13$, $\Omega /\omega _{0}=5$.

Figure 5. Second derivative of the Green function for the oscillator.
Parameters for the single relaxation time model are $\gamma /\omega
_{0}=10/13$ and $\Omega /\omega _{0}=5$.

Figure 6. Comparison of the exact and weak coupling expressions for the mean
square displacement at zero temperature for the oscillator. The parameters
chosen are $\gamma /\omega _{0}=10/13$ and $\Omega /\omega _{0}=5$.

\end{document}